\documentclass[a4paper,twocolumn,amsfonts,amssymb,amsmath]{revtex4}%
\usepackage{graphicx}
\usepackage{ae}
\usepackage[thinspace,thinqspace,amssymb,textstyle]{SIunits}

\usepackage{hyperref}
\usepackage[all]{hypcap}

\newcommand{\diff}[2]{\dfrac{ \textrm{d} #1}{\textrm{d} #2}}
\newcommand{\difff}[2]{\dfrac{ \textrm{d}^2 #1}{\textrm{d} #2^2}}

\newcommand{\tavg}[1]{\left \langle {#1} \right \rangle}
\newcommand{\bigo}[1]{\mathcal{O}\left(#1\right)}
\newcommand{\vect}[1]{\mathbf{#1}}
\newcommand{\tilchi}{\widetilde{\chi}}
\newcommand{\gammaf}[1]{\Gamma \left( #1 \right)}
\newcommand{\abs}[1]{\left| #1 \right|}
\newcommand{\Ai}{\text{Ai}}
\newcommand{\Bi}{\text{Bi}}
\newcommand{\GeV}{\giga\electronvolt}

\begin{document}

\title{Sommerfeld Enhancement of DM Annihilation:\\ Resonance Structure, Freeze-Out and CMB Spectral Bound}
\author{Steen Hannestad}
\author{Thomas Tram}
\email[Comments: ]{tram@phys.au.dk}
\affiliation{Department of Physics and Astronomy, University of Aarhus, 8000 Aarhus C, Denmark}

\date{\today}

\begin{abstract}
In the last few years there has been some interest in WIMP Dark Matter models featuring a velocity dependent cross section through the Sommerfeld enhancement mechanism, which is a non-relativistic effect due to massive bosons in the dark sector. In the first part of this article, we find analytic expressions for the boost factor for three different model potentials, the Coulomb potential, the spherical well and the spherical cone well and compare with the numerical solution of the Yukawa potential. We find that the resonance pattern of all the potentials can be cast into the same universal form.

In the second part of the article we perform a detailed computation of the Dark Matter relic density for models having Sommerfeld enhancement by solving the Boltzmann equation numerically. We calculate the expected distortions of the CMB blackbody spectrum from WIMP annihilations and compare these to the bounds set by FIRAS. We conclude that only a small part of the parameter space can be ruled out by the FIRAS observations.
\end{abstract}

\maketitle
\section{Introduction}
The most probable candidate for Dark Matter is arguably the WIMP, a Weakly Interacting Massive Particle. This is due to what is often called the 'WIMP miracle'; new physics is required at the TeV scale when unitarity breaks down in the Standard Model, and some symmetry should protect the proton from decaying too fast through new diagrams involving these new degrees of freedom. The symmetry may ensure that at least one of the new particles is stable on cosmological timescales, while the annihilation cross section is typically within an order of magnitude of what is required to explain the observed relic abundance.

In the last few years there have been some observations~\cite{Adriani:2008zr,Barwick:1997ig,Beatty:2004cy,Aguilar:2007yf,Chang:2008zzr,Finkbeiner:2003im,Dobler:2007wv,Strong:2005zx,Thompson:2004ez} pointing to anomalous emission within the galaxy. It is possible, that these observations may be explained by WIMPs annihilating in the galactic halo~\cite{Kane:2002nm,Hooper:2003ad,Cholis:2008qq,Baltz:2001ir,Hooper:2007kb}. Common to all these suggestions is that the necessary annihilation rate requires an annihilation cross section a few orders of magnitude larger than the one consistent with a thermal relic density, $\tavg{\sigma v}_\text{TH} \simeq 3\cdot 10^{-26}\centi\meter\cubed\per\second$.

This apparent discrepancy, as well as the necessary suppression of hadronic final states, has been suggested to stem from new GeV-mass force carriers in the dark sector~\cite{Cirelli:2008pk,ArkaniHamed:2008qn} which leads to Sommerfeld enhancement of the Dark Matter pair annihilation cross section. Although the anomalies mentioned above may turn out to be explained by normal astrophysical processes, there may still be force carriers in the dark sector which are light compared to the mass of the Dark Matter particle, so the mechanism of Sommerfeld enhancement is interesting and generic. Sommerfeld enhancement applied to WIMP annihilations in general, not related to the cosmic ray excess, was first studied in~\cite{Hisano:2003ec,Hisano:2004ds} and later in~\cite{Cirelli:2007xd} and~\cite{MarchRussell:2008yu}.
\section{Sommerfeld enhancement}
Sommerfeld enhancement is a consequence of having light force carriers mediate an attractive interaction in the dark sector, which creates a Yukawa-potential for the WIMPs. In this work we consider only s-wave annihilation. We write this potential as
\begin{align}
V(r) &= - \frac{\lambda^2}{4\pi r} e^{-m_\phi r} = -\frac{\alpha}{r} e^{-m_\phi r}, \label{eq:yukawa}
\end{align}
where $\lambda$ is the coupling parameter, $m_\phi$ is the mass of the force carrier and $r$ is the relative distance between the WIMPs. Since the potential is only a function of the relative distance, it only enters in the reduced one-particle Schrödinger equation for the relative motion. In spherical coordinates this is
\begin{align}\label{eq:fullscrodinger}
-\frac{1}{2\mu} \nabla ^2 \psi_k &=\left( \frac{k^2}{2\mu} - V(r) \right) \psi_k,
\end{align}
where $\mu = m_\chi/2$ is the reduced mass. We are interested in the probability density of $\psi_k$ when $r=0$, since the boost factor for the s-wave case is given by $S_k = |\psi_k(0)|^2$, when $\psi_k$ is normalized to the asymptotic form
\begin{align}
\psi &\rightarrow e^{i k z} + f(\theta) \frac{e^{ikr}}{r} \text{ as }r \rightarrow \infty.
\end{align}
This form of the boost factor can be derived~\cite{ArkaniHamed:2008qn} from a simple argument: Annihilations are assumed to proceed by a delta function interaction, so the rate of this process must be proportional to the norm squared of the reduced wave function at zero separation, $\sigma \propto \left|\psi(r=0)\right|^2$. Denoting the $V=0$ wave function by $\psi^0$, we must have
\begin{align}
S&=\frac{\sigma}{\sigma^0} = \frac{\left|\psi(0)\right|^2}{\left|\psi^0(0)\right|^2}.
\end{align}

When the potential is rotationally symmetric, we know that the solution of equation \eqref{eq:fullscrodinger} becomes invariant under rotations around the axis of the incoming particle, and hence we can expand $\psi$ in products of Legendre polynomials and a radial wave function $R_{kl}$. Only $R_{k0}$ is non-zero at the origin, so we define $\chi_k \equiv rR_{k0}$, which leads to the equation
\begin{align}
\frac{1}{m_\chi} \difff{\chi}{r} &=\left( -\frac{\alpha}{r} e^{-m_\phi r} - m_\chi \beta^2\right) \chi \Rightarrow  \label{eq:radeq1} \\
\frac{m_\phi}{m_\chi} \difff{\chi}{x} &= \left( -\frac{\alpha}{x} e^{-x} - \frac{m_\chi}{m_\phi}\beta^2 \right) \chi \Rightarrow  \label{eq:radeq2} \\
\difff{\chi}{x} &= - \left(\frac{\alpha}{f x} e^{-x} + \left(\frac{\beta}{f}\right)^2 \right) \chi \label{eq:radeq3},
\end{align}
where we introduced the dimensionless distance $x\equiv m_\phi r$ and the ratio of the two masses $f \equiv m_\phi / m_\chi$. Equation \eqref{eq:radeq3}, together with the boundary conditions
\begin{subequations} \label{eq:bvp}
\begin{align}
\chi(0) &= 0 \\
\chi &\rightarrow \sin \left(\frac{\beta}{f} x + \delta   \right),\text{ as } x \rightarrow
\infty. \label{eq:bvright}
\end{align}
\end{subequations}
defines the problem.

\subsection{Numerical solution}
\begin{figure}%
		\begin{center}				\includegraphics[height=1.0\columnwidth,angle=270,clip=true]{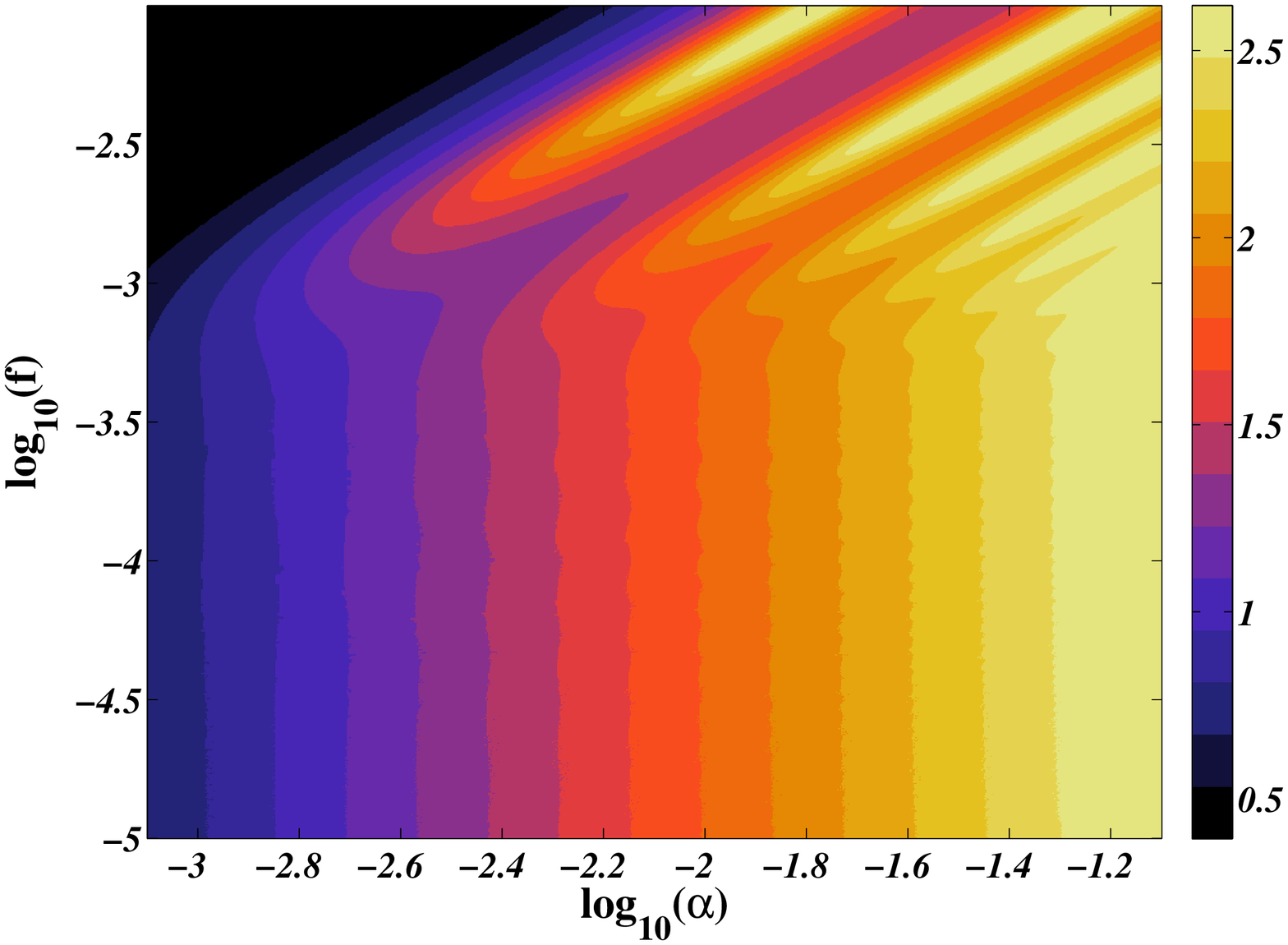}%
		\end{center}
		\caption{Sommerfeld boost factor in $\log_{10}$ for a relative velocity of $150\kilo\meter\per\second$.}%
		\label{fig:boost_8_200}%
	\end{figure}

In principle we should solve the boundary value problem \eqref{eq:bvp} numerically, but since the Schrödinger equation is linear, we can exchange the condition \eqref{eq:bvright} by setting the derivative of $\chi$ to unity at $x=0$ and solve what is now an initial value problem for $\widetilde{\chi}$. The solution to the original problem is just given by $\chi = \widetilde{\chi}/A$, where $A$ is the asymptotic amplitude of $\widetilde{\chi}$. The enhancement factor is then given by
\begin{align}
S_k &=\left| \frac{R_{k0}^2(0)}{k} \right|^2 = \left|\frac{1}{k} \diff{\chi_k}{r}(0) \right|^2 =
\left|\frac{f}{A \beta} \right|^2. \label{eq:boostcalc}
\end{align}
In figure~\ref{fig:boost_8_200} we have plotted the Sommerfeld boost factor for a velocity of $150\kilo\meter\per\second$ for the whole $(\alpha,f)$-parameter space. The vertical lines in figure~\ref{fig:boost_8_200} are ragged by thin, straight lines of resonance which have, as we shall see later, a rather large impact on the freeze out process.

To calculate the amplitude $A$ in \eqref{eq:boostcalc}, we must know when our waveform $\widetilde{\chi}$ has reached its asymptotic form. This must happen when the Yukawa potential becomes much less than the kinetic energy of the particle, so we find the position of equality:
\begin{align}
\frac{\alpha}{x_\text{range}}e^{-x_\text{range}} &= \frac{\beta^2}{f} \Rightarrow\nonumber \\
x_\text{range} e^{x_\text{range}} &= \frac{f \alpha}{\beta^2} \Rightarrow \nonumber\\
x_\text{range} &= W\left( \frac{f \alpha}{\beta^2} \right)\label{eq:xrange},
\end{align}
where $W$ is the Lambert W-function. To calculate the amplitude we use the asymptotic waveform \eqref{eq:bvright}:
\begin{subequations}
\begin{align}
\tilchi &=A \sin \left(\frac{\beta}{f}x+ \delta \right)\\
\diff{\tilchi}{x} &= \frac{\beta}{f} A \cos \left(\frac{\beta}{f}x+\delta \right),
\end{align}
\end{subequations}
which leads to this expression for the amplitude $A$:
\begin{align}
A &= \sqrt{\tilchi^2+\left(\frac{f}{\beta} \diff{\tilchi}{x} \right)^2}.
\end{align}
%
%\begin{align}
%\frac{\tilchi}{\diff{\tilchi}{x}} &=\frac{f}{\beta} \tan\left( \frac{\beta}{f} x + \delta %\right) \Rightarrow \\
%A &= \frac{\tilchi}{\sin \left( \arctan \left[ \left( \frac{f}{\beta \tilchi} \diff{\tilchi}{x} \right)^{-1} \right] \right)}.
%\end{align} 

We evolve our waveform $\tilchi$ to $1.5 x_\text{range}$, and after this point we calculate $A$ at each succeeding point until $A$ has converged. 

\subsection{Coulomb potential}
In the limit of a massless force carrier, the Yukawa potential becomes a Coulomb potential. In this case, the enhancement factor can be calculated in closed form,
\begin{equation}
S_{\epsilon_C} = \frac{\pi/\epsilon_C}{1-e^{-\pi/\epsilon_C}}, \qquad \epsilon_C=\frac{\beta}{\alpha} \label{eq:coulombboost}
\end{equation}
by solving the Schrödinger equation in terms of hyper geometric functions. The derivation of equation \eqref{eq:coulombboost} can be found in section \ref{sec:coulomb}.
We want to stress, that the applicability of this limit does not only depend on $f\ll 1$, but is also dependent on $\alpha$ and $\beta$. The correct question to ask is whether or not the scattering will take place in a regime in which the potential looks like a Coulomb potential. We want the range of the potential $x_\text{range}$ from \eqref{eq:xrange}, to be less than something like $\ln(2)$. This would keep the exponential in the Yukawa potential at order 1 during the interaction, so we should be safe in assuming a Coulomb potential instead. (However, this does not take into account the non-perturbative effects of resonances.)
\begin{figure}%
		\begin{center}				\includegraphics[height=1.0\columnwidth,angle=270,clip=true]{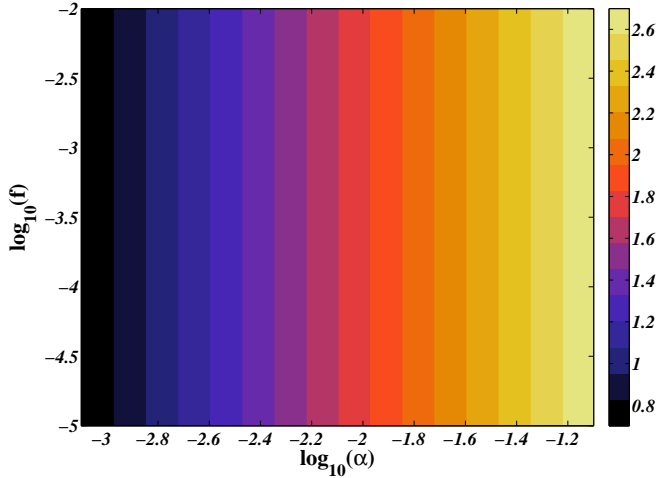}%
		\end{center}
		\caption{Boost factor in $\log_{10}$ for the Coulomb potential in $\log_{10}$ for a relative velocity of $150\kilo\meter\per\second$.}%
		\label{fig:coulombboost}%
\end{figure}
The boost factor \eqref{eq:coulombboost} is plotted in figure \ref{fig:coulombboost}. If we compare with the Sommerfeld-case in figure \ref{fig:boost_8_200}, we can see that the lower left part of the Sommerfeld parameter space is well described by the boost from the Coulomb potential. This part of the parameter space is also the part where the range \eqref{eq:xrange} of the potential is less than about $0.7$ as expected. However, we do not find resonances in the Coulomb case as we do for the Sommerfeld enhancement, because the potential is not localized. To learn more about these resonances, we relate the Yukawa potential to two model potentials which can be solved analytically.

\subsection{Spherical Well}
The resonance structure of the Sommerfeld-enhancement does not appear in the massless limit, but we can understand it qualitatively by examining the spherical well and relating the well to the Yukawa potential\footnote{In \cite{Hisano:2004ds}, the authors also related the Yukawa potential to the spherical well to understand the effect of resonances.}. We look at a potential of the form
\begin{equation}
V(r) = \left\{
\begin{array}{rl}
-V_0, & r \leq L/m_\phi\\
0,    & r > L/m_\phi.
\end{array} \right.
\end{equation}
We set $V_0$ to a value such that the potential integrated over volume agrees with the Yukawa potential \eqref{eq:yukawa}. We find
\begin{align}
V_0 &= \frac{3\alpha m_\phi}{L^3},
\end{align}
where $L$ is the range of the potential in units of $m_\phi^{-1}$, and should be order $1$. The Schrödinger equation for this potential is
\begin{align}
\difff{\chi}{x} &= - \left(\frac{V_0}{m_\phi f} + \left(\frac{\beta}{f}\right)^2 \right) \chi \Rightarrow \\
\difff{\chi}{x} &= - \left( K^2 + \epsilon^2 \right),
\end{align}
where we are using $x = m_\phi r$ and
\begin{align}
K &= \sqrt{\frac{3\alpha}{fL^3}} \quad \epsilon = \frac{\beta}{f}.
\end{align}
This equation is solved by sines and cosines, so if we define $p=\sqrt{K^2 + \epsilon^2}$, we can write the general solution as
\begin{subequations}\label{eq:chiwell}
\begin{align}
\chi^{x<L}(x) &= A \sin \left( px + \gamma \right) \label{eq:chiwellin}\\
\chi^{x>L}(x) &= B \sin \left( \epsilon(x-L)+ \delta \right) \label{eq:chiwellout},
\end{align}
\end{subequations}
and using boundary conditions \eqref{eq:bvp}, we can set $\gamma=0$ and $B=1$. We must now match the wave function and its derivative at the boundary at $x=L$. We get the equations
\begin{subequations}
\begin{align}
A \sin \left( pL \right) &= \sin \left( \delta \right) \\
Ap \cos \left(pL \right) &= \epsilon \cos \left( \delta \right)
\end{align}
\end{subequations}
and we solve for the amplitude $A$. We find:
\begin{align}
A^2 &= \frac{1}{1+\left(\frac{K}{\epsilon}  \right)^2 \cos^2\left(pL \right) },
\end{align}
which results in a boost factor
\begin{align} \label{eq:boostfromwell}
S &= \frac{1 + \left(\frac{K}{\epsilon}  \right)^2}{1+\left(\frac{K}{\epsilon}  \right)^2 \cos^2\left(pL \right) }.
\end{align}

\begin{figure}
\begin{center}
\includegraphics[height=1.0\columnwidth,angle=270,clip=true]{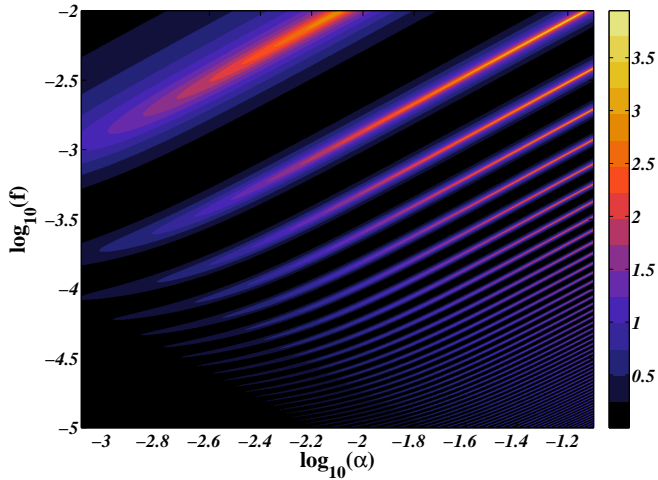}%
\end{center}
\caption{Boost factor in $\log_{10}$ for the spherical well with $v=150\kilo\meter\per\second$. The depth of the well is related to the strength of the corresponding Yukawa potential.} 
\label{fig:well1}%
\end{figure}
We have plotted this boost factor for $L=1$ in figure \ref{fig:well1} for the same $(\alpha,f)$-parameter space as before. If we compare with figure \ref{fig:boost_8_200}, we can see the same type of resonances, although they become suppressed in the Yukawa case when the scattering becomes Coulomb-like. The structure of \eqref{eq:boostfromwell} tells us something about the behavior of resonances, even for the Yukawa-case. First off, since the denominator is always larger than unity, the maximal boost must be of the order $\sim K^2/\epsilon^2$. Secondly, if the cosine is of order unity, the boost will be order unity as well. So the resonances occur when $pL/\pi \simeq n + 1/2$.

\begin{figure}
\begin{center}
\includegraphics[height=1.0\columnwidth,angle=270,clip=true]{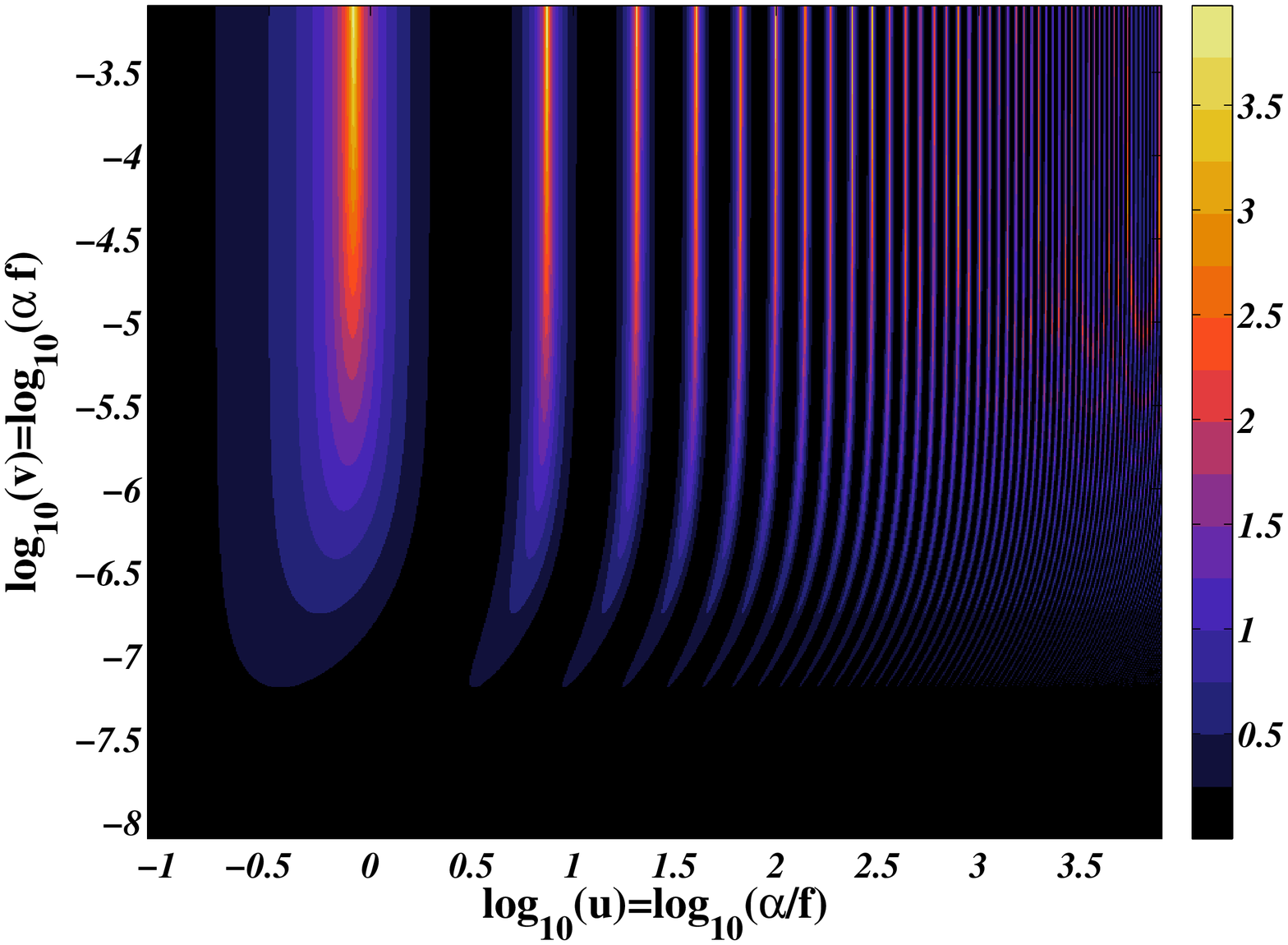}%
\end{center}
\caption{Boost factor for the spherical well with $v=150\kilo\meter\per\second$ in the new coordinates $u$ and $v$.} 
\label{fig:well2}%
\end{figure}
If we are interested in resonances giving $S$ of order $100$ or more, we can make the approximation $\epsilon/K \ll 1$, which allows us to use the zeroth order approximation $p \approx K$. Using the definition of $K$, we find
\begin{align}
\frac{KL}{\pi} &= \sqrt{\frac{3 \alpha}{f L^3 }}\frac{L}{\pi} =\sqrt{\frac{\alpha}{f}} \sqrt{\frac{3}{L \pi^2}}, \label{eq:KLpi}
\end{align}
so the position of the resonances depends only on the ratio of $\alpha$ and $f$. This suggests that we introduce the rotated coordinates
\begin{subequations}\label{eq:uvcoords}
\begin{align}
u &= \frac{\alpha}{f} \\
v & =\alpha f.
\end{align}
\end{subequations}
Using \eqref{eq:KLpi}, we find that the resonances happens at
\begin{align}
\sqrt{u_n} &= \sqrt{\frac{L \pi^2}{3}} \left(n + \frac{1}{2} \right) \Rightarrow \nonumber \\
u_n &= \frac{L\pi^2}{3} \left(n + \frac{1}{2} \right)^2,\quad n=0,1,\ldots \label{eq:unwell}
\end{align}
Since the distance between resonances increases only by $n^2$, it is clear that they will happen closer and closer when viewed in a logarithmic plot, just as we saw in the Yukawa-case. The other parameter $v$ controls the order of the boost factor, since we have
\begin{align}
\frac{K^2}{\epsilon^2} &= \frac{3}{L^3} \frac{f\alpha}{\beta^2} = \frac{3}{L^3} \frac{v}{\beta^2}.
\end{align}
\begin{figure}
\begin{center}
\includegraphics[height=1.0\columnwidth,angle=270,clip=true]{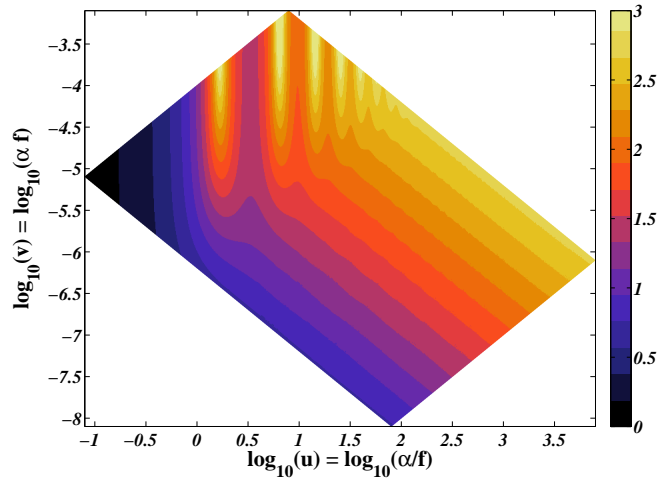}%
\end{center}
\caption{The Sommerfeld boost factor with $v=150\kilo\meter\per\second$ in the new coordinates $u$ and $v$. We have cropped the plot at the $(\alpha,f)$ limits from figure \ref{fig:boost_8_200}} 
\label{fig:uvsf}%
\end{figure}
We have shown the boost plot in the rotated coordinates $(u,v)$ in figure \ref{fig:well2}. We suspect that this choice of coordinates would also work well in the Yukawa case, so we calculate the Sommerfeld boost factor in the $(u,v)$ coordinates, and the result is shown in figure \ref{fig:uvsf}. As is clear from the plot, $u$ alone determines the position of the peaks, just as for the spherical well. This is in agreement with~\cite{Cirelli:2007xd}, as our coordinate $u$ is identical to their parameter $\epsilon_\phi$. 

\subsection{Spherical Slope Well}
We now examine another model potential, the spherical slope well, to see how the position of resonances change compared to the spherical well. We write the potential
\begin{equation}
V(r) = \left\{
\begin{array}{rl}
-V_0\left(1-\frac{m_\phi}{L} r\right), & r \leq L/m_\phi\\
0,    & r > L/m_\phi,
\end{array} \right.
\end{equation}
and as before, we set $V_0$ to a value such that the potential integrated over volume agrees with the Yukawa potential \eqref{eq:yukawa}. In this case we find
\begin{align}
V_0 &= \frac{12\alpha m_\phi}{L^3},
\end{align}
where $L$ is the range of the potential in units of $m_\phi^{-1}$, and should be order $1$. The Schrödinger equation for this potential is
\begin{align*}
\difff{\chi}{x} &= - \left(\frac{V_0}{m_\phi f}\frac{1}{L} \left(L-x \right) + \left(\frac{\beta}{f}\right)^2 \right) \chi \Rightarrow \\
\difff{\chi}{x} &= - \left( K^2 (L-x) + \epsilon^2 \right) \Rightarrow\\
\difff{\chi}{x} &= - \left( -K^2 x + p^2 \right),
\end{align*}
where we are using $x = m_\phi r$, $K = \sqrt{\frac{12\alpha}{fL^4}}$ and $p = \sqrt{K^2L+\epsilon^2}$. If we now do the substitution $\xi = K^{-4/3}\left(K^2x -p^2 \right)$, we get the Airy equation
\begin{align}\label{eq:airyeq}
\difff{\chi}{\xi} &=\xi \chi,
\end{align}
which has the general solution
\begin{align}
\chi &= c_1 \Ai(\xi) + c_2 \Bi(\xi).
\end{align}

We are interested in $\chi$ at two different values, $\xi_0 \equiv \xi(x=0)$ and $\xi_L \equiv \xi(x=L)$. These are given by
\begin{subequations}
\begin{align}
\xi_0 &= K^{-4/3}\left( -p^2 \right) =-K^{-4/3}p^2\\
\xi_L &= K^{-4/3}\left( K^2L-p^2 \right) = -K^{-4/3}\epsilon^2.
\end{align}
\end{subequations}
Since we have $\xi<0$ in both cases, it makes sense to introduce the variable
\begin{align}
\zeta = \frac{2}{3}\left(-\xi\right)^{3/2},
\end{align}
and reexpress the solution $\chi$ and its derivative in terms of Bessel functions. We have:
\begin{subequations}\label{eq:bessel_sol}
\begin{align}
\chi(\xi) &=\sqrt{-\xi}\left[-AJ_{1/3}(\zeta) +BJ_{-1/3}(\zeta) \right]\label{eq:bessel1} \\
\diff{\chi}{\xi}(\xi) &= -\xi \left[AJ_{-2/3}(\zeta) + BJ_{2/3}(\zeta) \right].\label{eq:bessel2} 
\end{align}
\end{subequations}
where the new coefficients $A$ and $B$ are related to the old ones by
\begin{equation}
A = -c_1/3+c_2/\sqrt{3} \qquad B = c_1/3+c_2/\sqrt{3}.
\end{equation}
Since we require $\chi(\xi_0) = 0$, equation \eqref{eq:bessel1} can be used to relate $A$ and $B$, since we have
\begin{align}
AJ_{1/3}(\zeta_0) &= BJ_{-1/3}(\zeta_0) \Rightarrow \nonumber \\
B &= \frac{J_{1/3}(\zeta_0)}{J_{-1/3}(\zeta_0)}A \equiv tA.
\end{align}
At this point we have one free parameter $A$ left to determine from the solution $\chi^{x<L}$. The solution outside the well is just as before, \eqref{eq:chiwellout}, and contains one free parameter, $\delta$. Both parameters are fixed by matching the solution and its derivative at the boundary; we have two equations with two unknowns:
\begin{subequations}
\begin{align}
\sin(\delta) &= \chi^{x<L}(x=L) \\
\epsilon \cos(\delta) &=\diff{\chi}{x}^{x<L}(x=L),
\end{align}
\end{subequations}
and the interesting parameter $A$ is most easily found by inserting the right hand sides in the well known trigonometric identity
\begin{align}
\epsilon^2 &= \epsilon^2 \sin^2(\delta) + (\epsilon \cos(\delta))^2 \nonumber \\
	&= \epsilon^2 (-\xi_L) A^2 \left[-J_{1/3}(\zeta_L) + tJ_{-1/3}(\zeta_L) \right]^2 + \nonumber \\
	   &+\xi_L^2K^{4/3}A^2 \left[J_{-2/3}(\zeta_L) + tJ_{2/3}(\zeta_L) \right]^2 \nonumber \Rightarrow \\
		A^2&=\frac{K^{4/3}/\epsilon^2}{  \begin{array}{l} \left[-J_{1/3}(\zeta_L) + tJ_{-1/3}(\zeta_L) \right]^2+ \\ +\left[ J_{-2/3}(\zeta_L)+tJ_{2/3}(\zeta_L)\right]^2  \end{array} }.
\end{align}
Using equation \eqref{eq:boostcalc}, we can find the boost factor $S$:
\begin{align}
S &= \frac{p^4}{\epsilon^2}K^{-4/3}A^2\left[J_{-2/3}(\zeta_0) + tJ_{2/3}(\zeta_0) \right]^2 \nonumber \Rightarrow \\
S &= \frac{p^4}{\epsilon^4} \frac{\left[J_{-2/3}(\zeta_0) + tJ_{2/3}(\zeta_0) \right]^2}{
	 	\begin{array}{l} \left[-J_{1/3}(\zeta_L) + tJ_{-1/3}(\zeta_L) \right]^2 +\\ +\left[ J_{-2/3}(\zeta_L)+tJ_{2/3}(\zeta_L)\right]^2 \end{array}}. \label{eq:slopeboost1}
\end{align}
\begin{figure}
\begin{center}
\includegraphics[height=1.0\columnwidth,angle=270,clip=true]{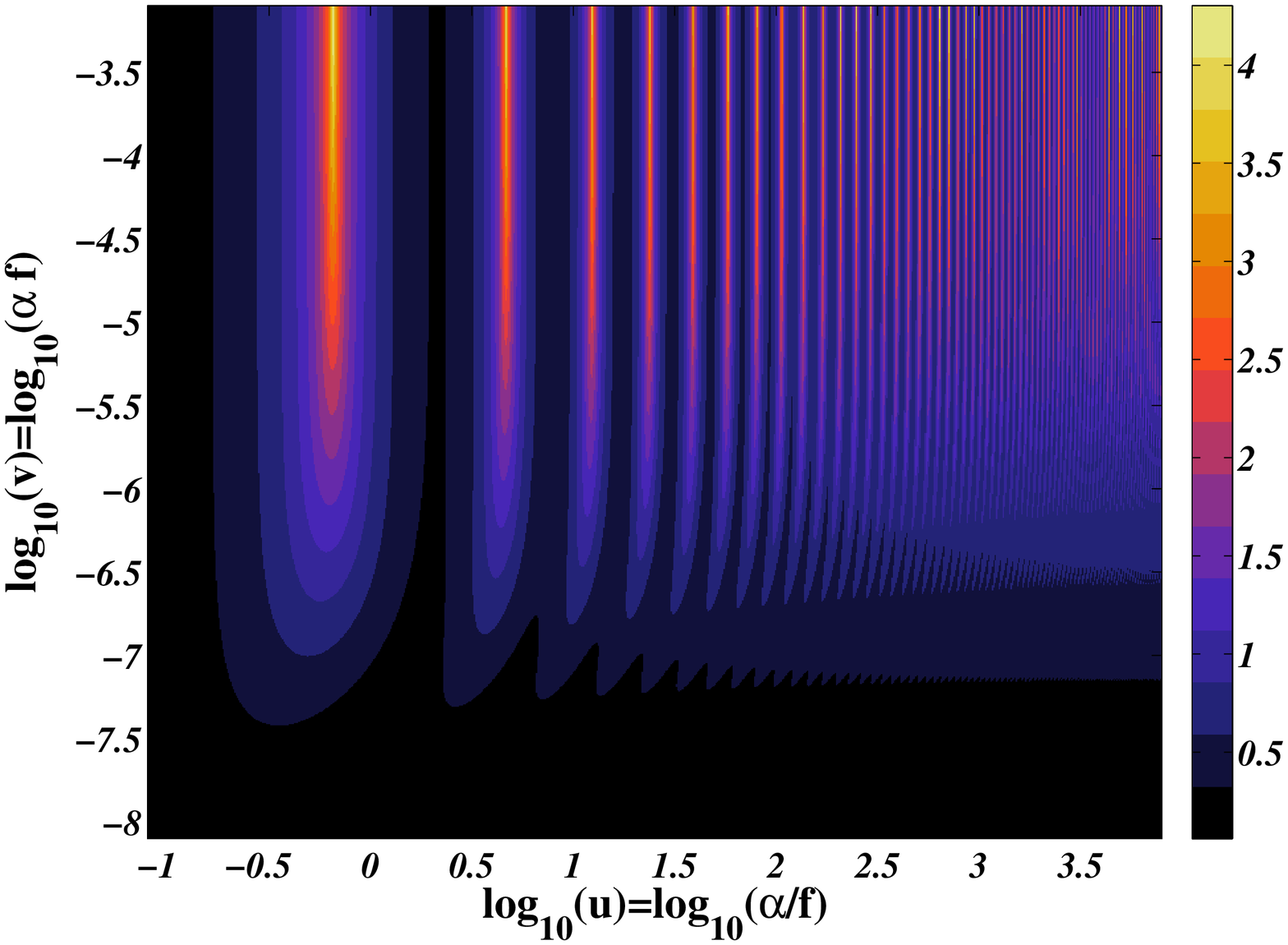}%
\end{center}
\caption{Boost factor in $\log_{10}$ for the spherical slope well with $v=150 \kilo\meter\per\second$. The depth of the well is related to the strength of the corresponding Yukawa potential.} 
\label{fig:slopewell1}%
\end{figure}
This is a rather complicated expression, but the numerator can be simplified by using the series expansion for $J(z)$ and the following identity for the gamma function:
\begin{subequations}
\begin{align}
\Gamma(1/3)\Gamma(2/3) &= \Gamma(1/3)\Gamma(1-1/3) \nonumber \\
                       &= \pi \csc (\pi/3) = 2\pi/\sqrt{3} \\
J_\nu(z)  &= (z/2)^\nu \sum_{k=0}^\infty{\frac{(-)^k (z/2)^{2k}}{k!\Gamma(\nu + k + 1)}}
\end{align}
\end{subequations}
Having these, we can prove the neat identity
\begin{align}
J_{-2/3}(z)J_{-1/3}(z) + J_{1/3}(z)J_{2/3}(z) &= \frac{\sqrt{3}}{\pi z}.
\end{align}
Reinserting $t$, the numerator of equation \eqref{eq:slopeboost1} can now be written as
\begin{align}
\text{num.} &= \left[J_{-2/3}(\zeta_0) + tJ_{2/3}(\zeta_0) \right]^2 \nonumber \\
						&= J_{-1/3}(z)^{-2} \times \nonumber \\
						&\times \left[J_{-1/3}(\zeta_0) J_{-2/3}(\zeta_0) +J_{1/3}(\zeta_0) J_{2/3}(\zeta_0) \right]^2 \nonumber \\
						&= \frac{3}{\pi^2 \zeta_0^2 J_{-1/3}(\zeta_0)^2}. \label{eq:nomsimple}
\end{align}
Inserting \eqref{eq:nomsimple} into \eqref{eq:slopeboost1} and rearranging the denominator, we find:
\begin{align}
S = &\frac{3^3 K^4}{2^2\pi^2 \epsilon^4 p^2} \bigg( J_{1/3}(\zeta_0)^2 \left[J_{-1/3}(\zeta_L)^2+J_{2/3}(\zeta_L)^2\right] + \nonumber \\
  &+J_{-1/3}(\zeta_0)^2 \left[J_{1/3}(\zeta_L)^2+J_{-2/3}(\zeta_L)^2\right] + \nonumber \\
  &+J_{-1/3}(\zeta_0)J_{1/3}(\zeta_0) J_{-1/3}(\zeta_L) \times \nonumber \\
  &\times \left[J_{-2/3}(\zeta_L)- J_{1/3}(\zeta_L) \right] \bigg)^{-1}. \label{eq:slopeboost2}
\end{align}
We have plotted this boost factor in figure \ref{fig:slopewell1} in the $(u,v)$-coordinates we introduced for the spherical well. We see the same behavior as before, but the expression for the boost factor is still too complicated for us to deduce where the resonances will be. Let us take another look at the denominator of equation \eqref{eq:slopeboost2}. We want to estimate the numerical value of the two arguments, $\zeta_0$ and $\zeta_L$. We got:
\begin{subequations}
\begin{align}
\zeta_0 &= \frac{2}{3}\left(-\xi_0 \right) ^{3/2} =\frac{2p^3}{3K^2} \\
\zeta_L &= \frac{2}{3}\left(-\xi_L \right) ^{3/2} =\frac{2\epsilon^3}{3K^2}.
\end{align}
\end{subequations}
If we are interested in regions of possibly large boost factors, we may assume $K \gg \epsilon$ as we did before, and we conclude that $\zeta_0 \gg 1$ and $\zeta_L \ll 1$. It turns out, that to a reasonable approximation, we can set the Bessel functions equal to their asymptotic values. If we remember that
\begin{subequations}
\begin{align*}
J_\nu(z) &\simeq \frac{\left(z/2\right)^\nu}{\Gamma(\nu + 1)}, &z\ll 1 \\
J_\nu(z) &\simeq \sqrt{\frac{2}{\pi z}} \cos\left(z - \nu \pi/2 -\pi/4 \right), &z\gg 1
\end{align*}
\end{subequations}
we can identify the two largest terms in the denominator of equation \eqref{eq:slopeboost2}. We are looking for the largest negative powers of $\zeta_L$, because they will give the largest contribution to the denominator. We have one term behaving as $\sim \zeta_L^{-4/3}$ and one term from the crossproduct is behaving as $\sim \zeta_L^{-1}$, and the rest of the 6 terms has more positive powers. These two worst terms multiply $J_{-1/3}(\zeta_0)$, so we suspect to have a resonance pattern that follow the zeros of $J_{-1/3}(\zeta_0)$, that is:
\begin{align}
J_{-1/3}(\zeta_0) &\sim \cos\left(\frac{2p^3}{3K^2} -\frac{1}{12}\pi \right)= 0 \Rightarrow \nonumber\\
\frac{2p^3}{3K^2} -\frac{1}{12}\pi &\simeq \frac{2K^3L^{3/2}}{3K^2} -\frac{1}{12}\pi = \pi \left(n + \frac{1}{2} \right) \Rightarrow \nonumber \\
\left(\frac{2}{3} K L^{3/2}    \right)^2 &= \left(n+\frac{7}{12} \pi\right) \Rightarrow \nonumber \\
u_n &=\frac{3 \pi^2}{16} L \left( n+ \frac{7}{12} \right)^2. \label{eq:unslope}
\end{align}

Thus, we have derived an analytic equation for the position of resonances, just as we did for the spherical well, and this expression agrees exactly with the resonances in figure \ref{fig:slopewell1}. 

\subsection{Hulthén potential}
Comparing equation~\eqref{eq:unwell} and~\eqref{eq:unslope}, we notice that they look very similar. Only the fraction in front and the 'phase' is different. It seems likely that we may fit the peaks of the resonances in the Sommerfeld case by an expression:
\begin{align}
u_n &= L\pi^2 \left(n + b\right)^2,
\end{align}
and this is indeed the case. We find the values $L=0.1592$ and $b=1.006$. The value of $b$ is very close to $1$, hinting that a similar treatment is doable in the Yukawa case, yielding a boost factor which depends only on a sine to lowest order. A model potential called the Hulthén potential admits an analytic solution for the s-wave case\footnote{We wish to thank the referee  for bringing this to our attention.}, and the Sommerfeld boost coming from this potential was studied in~\cite{Cassel:2009wt} and later in~\cite{Slatyer:2009vg}.

The potential looks like
\begin{align}
V_\text{H} = -\frac{A \delta e^{-\delta r}}{1-e^{-\delta r}},
\end{align}
where $A$ and $\delta$ are parameters. We stress that this is not a general version of the Yukawa potential, but it is a model potential just like the spherical well and the spherical slope well. However, unlike these two potentials, the Hulthén potential reproduces the $1/r$-behavior of the Yukawa potential in the limit $r\rightarrow 0$, and it decays exponentially instead of having a fixed range. By a procedure similar\footnote{\cite{Cassel:2009wt} argues using the Lippmann-Schwinger equation, that the first moments of the potentials should be set equal. In our approach we set the volumes equal instead, i.e. the second moment. We checked how using the first moment would affect our results, and the effect was only to change the factor in front of equation~\eqref{eq:unwell} and~\eqref{eq:unslope} as well as making them independent of $L$.} to ours, $A$ and $\delta$ is found to be $A=\alpha$ and $\delta = k m_\phi = \pi^2/6$. In our notation, the $l=0$ case of the Sommerfeld boost just before equation 44 in~\cite{Cassel:2009wt}, can be written as
\begin{align}
S_\text{H} &= \frac{\frac{\pi \alpha}{\beta} \sinh \left( 2\pi \frac{\beta}{k f} \right) }
	{\cosh \left( 2\pi \frac{\beta}{k f} \right) - 
	\cos \left( 2 \pi \frac{\beta}{k f} \sqrt{\frac{k \alpha f}{\beta^2} -1} \right)}.
	\label{eq:S_hulthen}
\end{align}
\begin{figure}%
	\begin{center}	\includegraphics[height=1.0\columnwidth,angle=270,clip=true]{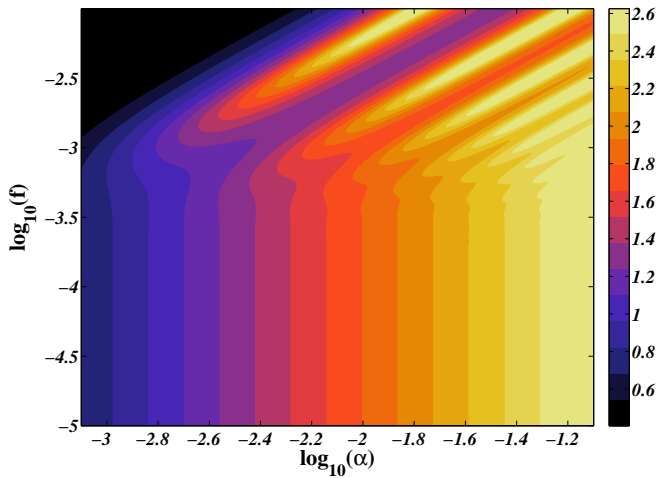}%
	\end{center}
\caption{Sommerfeld boost factor in $\log_{10}$ for the Hulthén potential at a relative velocity of $150\kilo\meter\per\second$.}%
\label{fig:S_hulthen}%
\end{figure}
Equation~\eqref{eq:S_hulthen} can be derived easily under the assumption $k \alpha f \leq \beta^2$, (which is usually not the case), however, a more careful derivation reveals this equation to be true in any case. We have plotted the Hulthén boost factor in figure~\ref{fig:S_hulthen}, and it looks very similar to the Yukawa case in figure~\ref{fig:boost_8_200}. Inspection of equation~\eqref{eq:S_hulthen} reveals the resonance pattern to be
\begin{align}
u_n 	&= k \left(n + 1 \right)^2 \nonumber \\
	&= \frac{\pi^2}{6} \left(n + 1 \right)^2. \label{eq:hulthen}
\end{align}

However, for large $n$, the resonance pattern is shifted completely because of the slight difference between $\pi^2/6$ and the fit value. For a velocity of $150\kilo\meter\per\second$ we found equation~\eqref{eq:S_hulthen} to be within 10\% of the numerical solution at 82\% of the plotted parameter space and within 30\% at 98\% of the parameter space. For a velocity of $10 \kilo\meter\per\second$, the Hulthén boost factor is within 10\% in only 43\% of the parameter space and within 30\% in 74\% of the parameter space. By using $k=0.1592\pi^2$ and simultaneously correcting the phase by $0.006$ by hand in equation~\eqref{eq:S_hulthen}, according to the fit parameters, we did somewhat better. The modified formula reproduced the resonance pattern perfectly, but the agreement of the boost magnitude at the first peak became worse.

\section{Constraining Sommerfeld Enhancement with the CMB}
There has been some work on constraining models incorporating Sommerfeld Enhancement using the CMB. WIMPs annihilating after recombination may contribute to reionisation, thereby changing the predicted optical depth which can be inferred from the CMB anisotropy spectrum. This was studied by \cite{Cirelli:2009bb}. Earlier, during the recombination phase, WIMP annihilations with Sommerfeld Enhancement also modify the standard scenario. An upper bound on this effect can again be inferred from the anisotropy spectrum, as was done in \cite{Chen:2003gz,Pierpaoli:2003rz,Furlanetto:2006wp,Mapelli:2006ej,Zhang:2007zzh,Galli:2009zc,Slatyer:2009yq}. We will focus on WIMP annihilations happening even earlier, in the redshift range of approximately $1100 < z < 2.1 \cdot 10^6$. Annihilations occurring in this redshift range will not influence the anisotropy spectrum, but they will distort the Black Body spectrum. 

Common to the above mentioned bounds is the fact that they use the expected cross section for a thermal relic, $\tavg{\sigma v}_\text{TH} \simeq 3\cdot 10^{-26}\centi\meter\cubed\per\second$, and multiply this with the boost factor to obtain the effective annihilation cross section. As was pointed out in~\cite{Dent:2009bv}, this is not strictly correct, since the freeze out process is modified by the boost mechanism. So to consistently probe the parameter space, we must first find the relativistic annihilation cross section $\sigma_0$, which gives the correct relic abundance.

\subsection{Relic density calculation} \label{sec:thermal_freezeout}
To catch the full nature of thermal freeze out in models with Sommerfeld enhancement, we must do a full calculation of the integrated Boltzmann equation because of the non-trivial velocity dependence. It should be noted, that we do expect something of the same order of magnitude as $\tavg{\sigma v}_\text{TH}$, since the Sommerfeld boost factor is very close to 1 at the time of freeze out. We start from the integrated Boltzmann equation for Dark Matter,
\begin{align}
a^{-3} \diff{(n_\chi a^3)}{t} &= \tavg{\sigma_\text{ann}v} \left\{ n_{\chi,\text{eq}}^2-n_\chi^2 \right\}, \label{eq:boltzmann1}
\end{align}
where $a$ is the scale factor, $n_\chi$ is the number density of Dark Matter, $n_{\chi,\text{eq}}$ is the Dark Matter equilibrium number density and $\tavg{\sigma_\text{ann}v}$ is the thermally averaged annihilation cross section. We normalize the number density to the total entropy density $s\propto a^{-3}$ by introducing $Y\equiv \frac{n_\chi}{s}$:
\begin{equation}
\diff{Y}{t} = s \tavg{\sigma_\text{ann} v} \left\{ Y_\text{eq}^2 - Y^2 \right\} \label{eq:boltzmann2}
\end{equation}
We also want to substitute the time parameter $t$ by the dimensionless evolution parameter $x\equiv \frac{m_\chi}{T_\gamma}$. After some manipulations, we find the final form of the Boltzmann equation:
\begin{align}
\diff{Y}{x} &= \sqrt{\frac{\pi}{45}} \frac{m_p m_\chi}{x^2}h_*^\frac{1}{2}\tavg{\sigma_\text{ann} v} \left\{ Y_\text{eq}^2 - Y^2 \right\}. \label{eq:evoboltz}
\end{align}
The manipulations leading up to equation \eqref{eq:evoboltz} can be found in section \ref{sec:boltzmann} of the appendix. In our case, $\sigma =\sigma_0 S(v,\ldots)$, where $\sigma_0$ is the s-wave annihilation cross section which is independent of velocity, and $S$ is the Sommerfeld boost factor, which depends on velocity and the model parameters. As was noted by the authors of~\cite{Feng:2009hw,Feng:2010zp}, $\sigma_0$ is not a completely free parameter but is related to $\alpha$ and $m_\chi$ by dimensional analysis, $\sigma_0 \sim \frac{\alpha^2}{m_\chi^2}$. However, the exact factor is highly model dependent, and for that reason we keep $\sigma_0$ as a parameter. 

Inserting the Sommerfeld cross section in equation \eqref{eq:evoboltz} leads to our final equation:
\begin{align}
\diff{Y}{x} &=\frac{m_p m_\chi}{x^2}h_*^\frac{1}{2} \sigma_0 \tavg{S(v,\ldots) v} 
		\left\{ Y_\text{eq}^2 - Y^2\right\}\label{eq:boltzmann25}\\
						&\equiv \lambda(x) \left\{ Y_\text{eq}^2 - Y^2 \right\}. \label{eq:boltzmann3}
\end{align}
The parameter $\sigma_0$ is then found by imposing the boundary condition that the WIMP must make up all of Dark Matter. In this article we are considering the case where the WIMP is not its own antiparticle, so we impose the boundary condition $\Omega_\text{DM} = 2\Omega_{\chi,0}$ assuming that there is no asymmetry between $\chi$ and its antiparticle.

The equilibrium number density $Y_\text{eq}$ can be calculated exactly under the assumption that Dark Matter follows a Maxwell-Boltzmann distribution. We got
\begin{equation}
Y_\text{eq}	= g_i \frac{45}{4\pi^4} \frac{x^2}{g_{*S}} K_2(x), \label{eq:Yeq}
\end{equation}
where $K_2$ is the modified Bessel function of the second kind. The derivation of equation \eqref{eq:Yeq} can be found in section \ref{sec:boltzmann} of the appendix. The values of $g_{*S}$ in equation~\eqref{eq:Yeq} and $h_*^\frac{1}{2}$ in equation~\eqref{eq:boltzmann25} depends on the temperature, and it is found by interpolation in a precomputed table. The number of internal degrees of freedom, $g_i$, was set to $2$. We also need the thermally averaged cross section. If we approximate the distribution function for the Dark Matter gas by the non-relativistic Maxwell-Boltzmann distribution
\begin{align}
f(\beta) &=\sqrt{\frac{2}{\pi}} x^\frac{3}{2} \beta^2 e^{-\frac{1}{2}x\beta^2}, 
\end{align}
the thermal average of $S$ becomes
\begin{align}
\tavg{S(\beta,\ldots) v} &\simeq \sqrt{\frac{2}{\pi}} x^\frac{3}{2} 
													\int_0^1 S(\beta,\ldots) \beta^2 e^{-\frac{1}{2}x\beta^2}. \label{eq:maxboltz}
\end{align}
This distribution is only normalized to $1$ when the upper limit goes to infinity, but if we are in the non-relativistic limit this is a small correction. (In the calculation we set the upper limit dynamically to 4 times the position of the peak of the distribution.) Before freeze out the gas is not strictly non-relativistic and we should instead use the relativistic distribution:
\begin{align}
f(\gamma)\mathrm{d}\gamma &= \frac{x}{K_2(x)} \beta \gamma^2(\beta) 				 	                           e^{-x\gamma(\beta)} \label{eq:maxjut1} \Rightarrow\\
f(\beta)\mathrm{d}\beta &= \frac{x}{K_2(x)} \gamma^5\beta^2 e^{-\gamma x}, \label{eq:maxjut2}
\end{align}
where $\gamma$ is the usual gamma factor and $K_2$ is the modified Bessel-function of the second kind. The thermal average then becomes
\begin{align}
\tavg{S v} &=\frac{x}{k_2(x)} \int_0^1 S(\beta,\ldots) \gamma^5\beta^2 e^{-x(\gamma-1)}, \label{eq:tavgrel}
\end{align}
where $k_2(x)=K_2(x)e^{x}$. When $\gamma$ becomes close to $1$ we must use the series expansion to calculate $1-\gamma$ for numerical stability. The difference in the required cross section $\sigma_0$ from using \eqref{eq:tavgrel} compared to \eqref{eq:maxboltz} was negligible, however. 

The numerical solution of equation~\eqref{eq:boltzmann3} is not entirely trivial, especially not since the equation is stiff and we need to solve it repeatedly. In section \ref{sec:boltzmann} of the appendix, we have described the numerical scheme we use for this problem, which is the same used in the DarkSUSY software package~\cite{DarkSUSY}. We recommend this scheme to others interested in Freeze-Out calculations.
\subsection{Distorting the Black Body spectrum}
When energy is injected into the CMB photons, two types of processes are needed to restore a black body spectrum: Number changing processes and equilibrating processes. Double Compton scattering and bremsstrahlung belongs to the first category, while Compton scattering and inverse Compton scattering belongs to the second. Double Compton scattering freezes out at $z_\text{DC} \simeq 2.1 \cdot 10^6$ and Compton scattering freezes out at $z_\text{C}\simeq 5.4 \cdot 10^4$. Energy which is deposited in the photon gas after $z_\text{DC}$ but before $z_\text{C}$ will be redistributed to give an entropy maximising Planck spectrum with a chemical potential $\mu$. Energy input after $z_\text{C}$, (but before recombination), can not equilibrate and will result in a Compton-y distortion~\cite{Bernstein:1989uq} of the Planck spectrum. 

The relevant quantity for deriving the size of both effects is thus the relative energy input to the CMB during these two epochs. Following~\cite{McDonald:2000bk}, we write
\begin{align}
\frac{\delta \rho_\gamma}{\rho_\gamma} &=
			\int_{t_1}^{t_2} \frac{\dot{\rho_\text{ann}}}{\rho_\gamma} \mathrm{d}t \nonumber \\
		&=\int_{t_1}^{t_2} \frac{2F m_\chi \tavg{\sigma_\text{ann} v} n_\chi^2}{\rho_{\gamma,0}
							 (1+z)^4} \mathrm{d}{t}, \label{eq:deltagamma}
\end{align}
where $\rho_\gamma$ is the energy density of the CMB photons and $z$ is the redshift. $F$ denotes the fraction of energy which is transfered to the CMB photons. This is independent of redshift when $z \gtrsim 2500$, and according to table 1 of~\cite{Slatyer:2009yq}, this is between 30\% and 90\% depending on the annihilation channel. The factor of $2$ in equation~\eqref{eq:deltagamma} stems from the fact that we, in consistence with our freeze out calculation, assume that $\chi$ is not its own antiparticle.

We introduce the following relations:
\begin{subequations} \label{eq:inputgamma}
\begin{align}
H^2(t) 	&= \frac{4 \pi ^3}{45 m_p^2} g_* T_{\gamma,0}^4 (1+z)^4 \Rightarrow  \label{eq:friedmann2}\\
\diff{t}{z} &= -\sqrt{\frac{45}{\pi}} \frac{m_p}{2\pi} g_*^{-\frac{1}{2}} T_{\gamma,0}^{-2}
								 (1+z)^{-3} \\
n_\chi 	&= n_{\chi,0} (1+z)^3 = \frac{\rho_{\chi,0}}{m_\chi} (1+z)^3  \nonumber\\
				&= \Omega_{\chi} \frac{3 H_0^2 m_p^2}{8 \pi m_\chi} (1+z)^3\\ 
\rho_\gamma &=\rho_{\gamma,0} (1+z)^4 \\
T_\gamma &=T_{\gamma,0} (1+z) \\
\tavg{\sigma_\text{ann} v} &=\sigma_0 \tavg{S(\beta,\ldots) v}
\end{align}
\end{subequations}
Using equations \eqref{eq:inputgamma} in equation \eqref{eq:deltagamma} yields:
\begin{align}
\frac{\delta \rho_\gamma}{\rho_\gamma} &=
		\frac{405}{64 \pi^5} \sqrt{\frac{5}{\pi}} g_*^{-\frac{1}{2}} F
  		\frac{\sigma_0}{m_\chi} \Omega_\chi^2 \left[ \frac{m_p^5 H_0^4}{T_{\gamma,0}^6} \right] \times \nonumber \\
  		&\times	 \int_{z(t_2)}^{z(t_1)} S_\text{avg}(z,\ldots) \frac{1}{1+z} \mathrm{d}z \nonumber \\
		&=\frac{405}{64 \pi^5} \sqrt{\frac{5}{\pi}} g_*^{-\frac{1}{2}} f
		   \left[\frac{100 \text{GeV}}{m_\chi} \right] \left[ \frac{\sigma_0}{10^{-26}	 \centi\meter\cubed\per\second} \right] \times \nonumber \\
		 &\times \left( \Omega_\chi h \right)^2 C_{-7} \int_{z(t_2)}^{z(t_1)} S_\text{avg}(z,\ldots) \frac{1}{1+z} \mathrm{d}z. \label{eq:deltagamma2}
\end{align}
Here $C_{-7}\simeq 2.8696\cdot 10^{-7}$ is a numerical constant. Since the Dark Matter is very cold at this time, the distribution function is strongly peaked around its mean value, so to a good approximation, we may write $S_\text{avg}(z) \simeq S(\beta_\text{mean}(z))$. We may also assume that the Sommerfeld enhancement has saturated at this stage, making the approximation $S(\beta_\text{mean}(z)) \simeq S(\beta_\text{mean}(z(t_1)))$, allowing us to pull $S$ outside the integral which can then be done analytically:
\begin{align}
\frac{\delta \rho_\gamma}{\rho_\gamma} &\approx 
 \frac{405}{64 \pi^5} \sqrt{\frac{5}{\pi}} g_*^{-\frac{1}{2}} F
		   \left[\frac{100 \text{GeV}}{m_\chi} \right] \left[ \frac{\sigma_0}{10^{-26} 
		   	 \text{cm}^3/s} \right]  \times \nonumber \\
		&\left( \Omega_\chi h \right)^2 C_{-7} S(\beta_\text{mean}(z_1)) \ln \left|\frac{1+z(t_1)}{1+z(t_2)} 			 						\right|.  
\end{align}
In these calculations we have assumed a radiation dominated universe as described by the Friedmann equation \eqref{eq:friedmann2}, which breaks down at $z\sim z_\text{eq}\simeq{3300}$, well before recombination. But as can be seen from the analytical approximation, the dependence on $z(t_2)$ is logarithmic, so the error in doing this is insignificant. When plotting the $\mu$ and $|y|$-distortions, we have put $F=1$ for convenience, but since the dependence is linear in $F$, it can be reinstated by multiplying the distortions by $F$.
\subsection{Results}
\begin{figure}
\begin{center}%
\includegraphics[height=1.0\columnwidth,angle=270,clip=true]{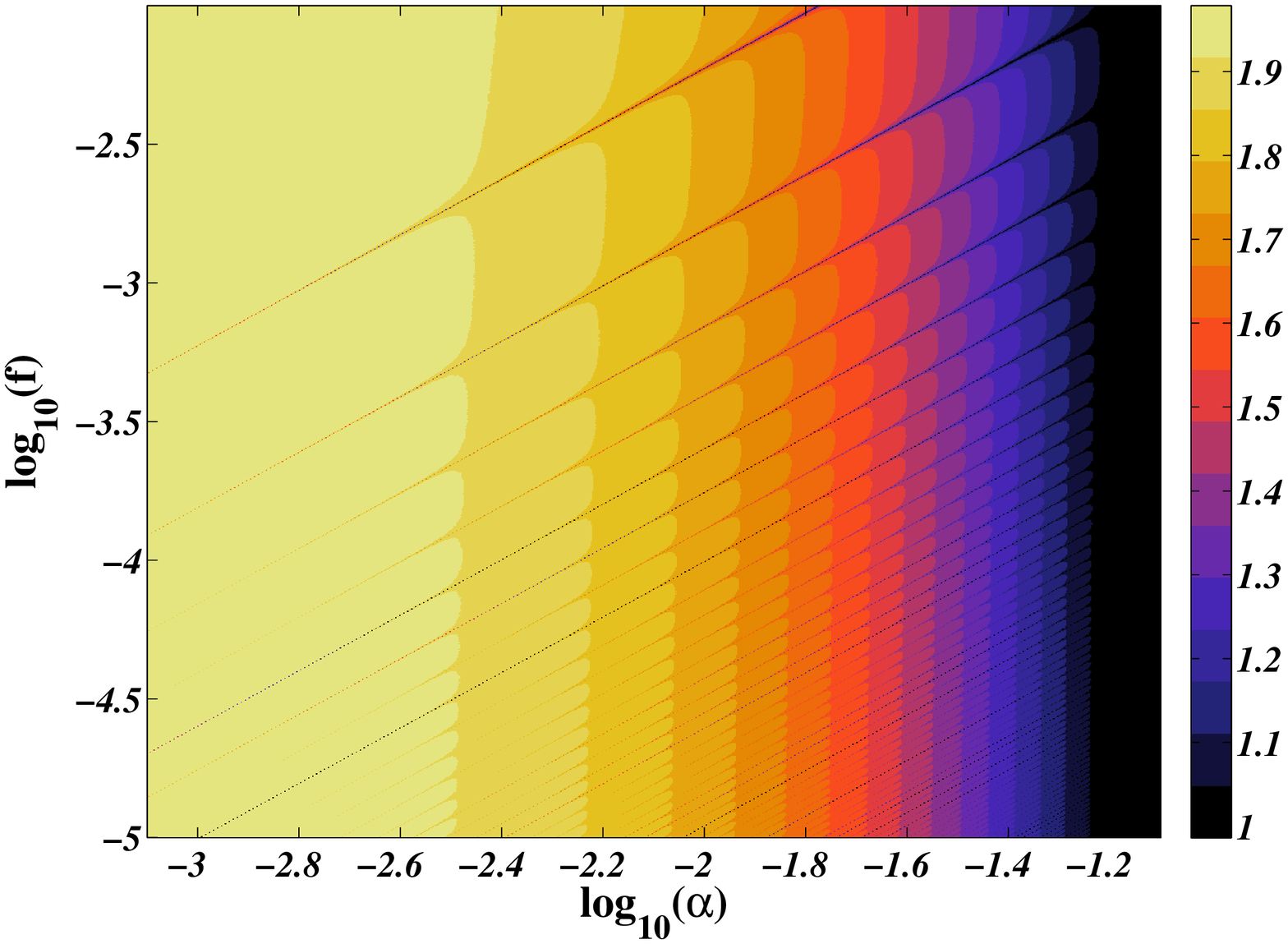}%
\end{center}
\caption{Required annihilation cross section $\sigma_0$ in units of $10^{-26}\centi\meter\cubed\per\second$ to explain the total Dark Matter abundance for a $200\giga\electronvolt$ WIMP and a kinetic decoupling temperature of $8\mega\electronvolt$.}%
\label{fig:sigma_8_200}%
\end{figure}

\begin{figure}
\begin{center}%
\includegraphics[height=1.0\columnwidth,angle=270,clip=true]{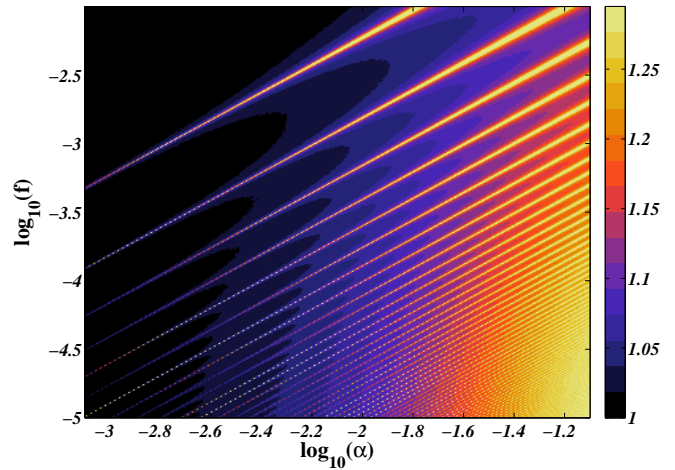}%
\end{center}
\caption{The ratio $\sigma_0^{\text{KD}=8\text{MeV}}/\sigma_0^{\text{KD}=500\text{MeV}}$ for a $200\text{GeV}$ WIMP.}%
\label{fig:divsigma_8_500}%
\end{figure}

\begin{figure}
\begin{center}%
\includegraphics[height=1.0\columnwidth,angle=270,clip=true]{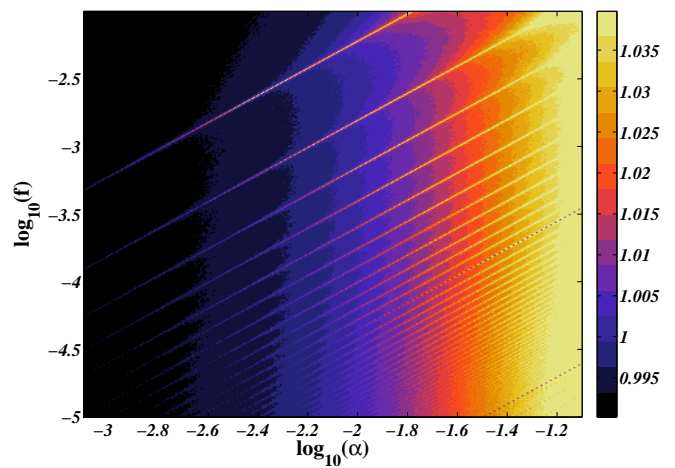}%
\end{center}
\caption{The ratio $\sigma_0/\sigma_0'$ of cross sections for a $200\GeV$ WIMP and a $1000\GeV$ WIMP respectively, which decouples kinetically at $x_\text{KD} = 2\cdot 10^3$.}%
\label{fig:divsigma_200_1000}%
\end{figure}

For each point in the $(\alpha,f)$-parameter space, we solved the Boltzmann equation from $x_1 = 1$ to $x_2 = m_\chi/T_{\gamma,0}$ to find the value of $\sigma_0$. This depends weakly on the WIMP mass as well as the kinetic decoupling temperature but the overall dependence on the Sommerfeld parameters can be seen in figure \ref{fig:sigma_8_200}.
The kinetic decoupling temperature was taken as a parameter, and the effect on the freeze out process is shown on figure \ref{fig:divsigma_8_500}. We find that this effect can be as large as 30\% in agreement with~\cite{Dent:2009bv,Zavala:2009mi}.
We expect $\sigma_0$ to be nearly independent of the WIMP mass and this is confirmed by figure~\ref{fig:divsigma_200_1000} which shows maximally one percent difference between a $200$GeV WIMP and a $1000\GeV$ WIMP for a kinetic decoupling value of $x_\text{KD} = 2\cdot 10^3$ for both. However, in a real particle physics model we expect $x_\text{KD}$ to have a slight mass dependence, so the actual effect of different masses may be somewhat bigger\cite{Bringmann:2009vf}.

\begin{figure}
\begin{center}%
\includegraphics[width=1.0\columnwidth,clip=true]{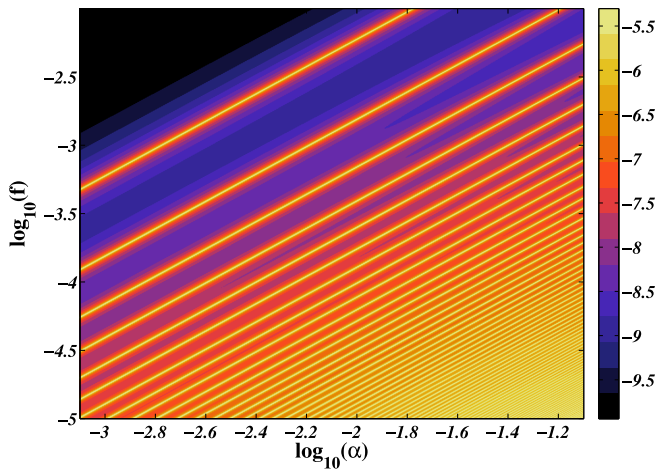}%
\end{center}
\caption{The magnitude of the Compton-$|y|$ parameter in $\log_{10}$ for a $200\text{GeV}$ WIMP and a kinetic decoupling temperature of $8\text{MeV}$.}%
\label{fig:cy_8_200}%
\end{figure}

\begin{figure}
\begin{center}%
\includegraphics[width=1.0\columnwidth,clip=true]{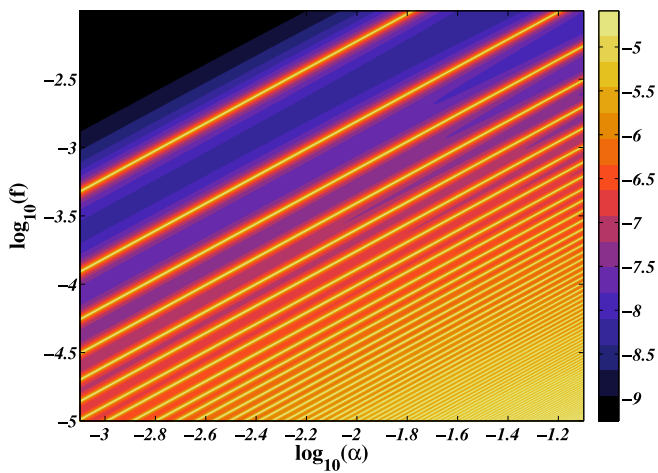}%
\end{center}
\caption{The magnitude of the chemical potential $\mu$ in $\log_{10}$ for a $200\text{GeV}$ WIMP and a kinetic decoupling temperature of $8\text{MeV}$.}%
\label{fig:mu_8_200}%
\end{figure}

For each point in the parameter space, we solve the integral in equation \eqref{eq:deltagamma2} on both the interval which is relevant for Compton-$|y|$ distortions as well as the interval relevant for the CMB photons to develop a chemical potential $\mu$. The results for a $200\GeV$ particle are shown in figure \ref{fig:cy_8_200} and \ref{fig:mu_8_200}. Considering the current bounds on $\mu$- and $y$-distortions from FIRAS \cite{Fixsen:1996nj} of $|\mu|<9\cdot 10^{-5}$, $|y|<1.5 \cdot 10^{-5}$, only a small portion of the parameter space can be ruled out. But there is a rather large part of lower right part of the parameter space which is close at saturating the current bound. The analysis for the $\mu$-distortion was already carried out in~\cite{Zavala:2009mi}, and our results agrees with theirs. 


As is evident from figure \ref{fig:cy_8_200} and \ref{fig:mu_8_200}, the two bounds are degenerate in the way that they both tends to rule out the resonances and the lower right part of the parameter space. With the current limits, the $|y|$-bound is always as strong or stronger than the $\mu$-bound. One would suspect this to be the case, since the $y$-distortion happens at a later time than the $\mu$-distortion, giving the WIMPs more time to cool. This is indeed true, but only for a small subset of the parameter space, at the resonances in the lower left part. For the rest of the parameter space, the Sommerfeld enhancement has already saturated at this point, and no further enhancement is possible.

%\begin{figure}
%\begin{center}
%\includegraphics[height=1.0\columnwidth,angle=270,clip=true]{ruledout_200_008_NT.eps}%
%\end{center}
%\caption{Green points are ruled out based only on the $\mu$-bound, red points are ruled out based only on the $y$-bound and yellow points are ruled out by both.}%
%\label{fig:ruledout_8_200}%
%\end{figure}
%\begin{figure}
%\begin{center}
%\includegraphics[height=1.0\columnwidth,angle=270,clip=true]{divboost_z1_z2_NT.eps}%
%\end{center}
%\caption{The ratio of the boostfactor at $z_\text{DC}$ and $z_\text{C}$ in $\log_{10}$. $z_\text{DC}$ is the time when the $\mu$-distortion starts happpening and $z_\text{C}$ when the $y$-distortion starts happening.}%
%\label{fig:divboost}%
%\end{figure}

\begin{figure}
\begin{center}
\includegraphics[height=1.0\columnwidth,angle=270,clip=true]{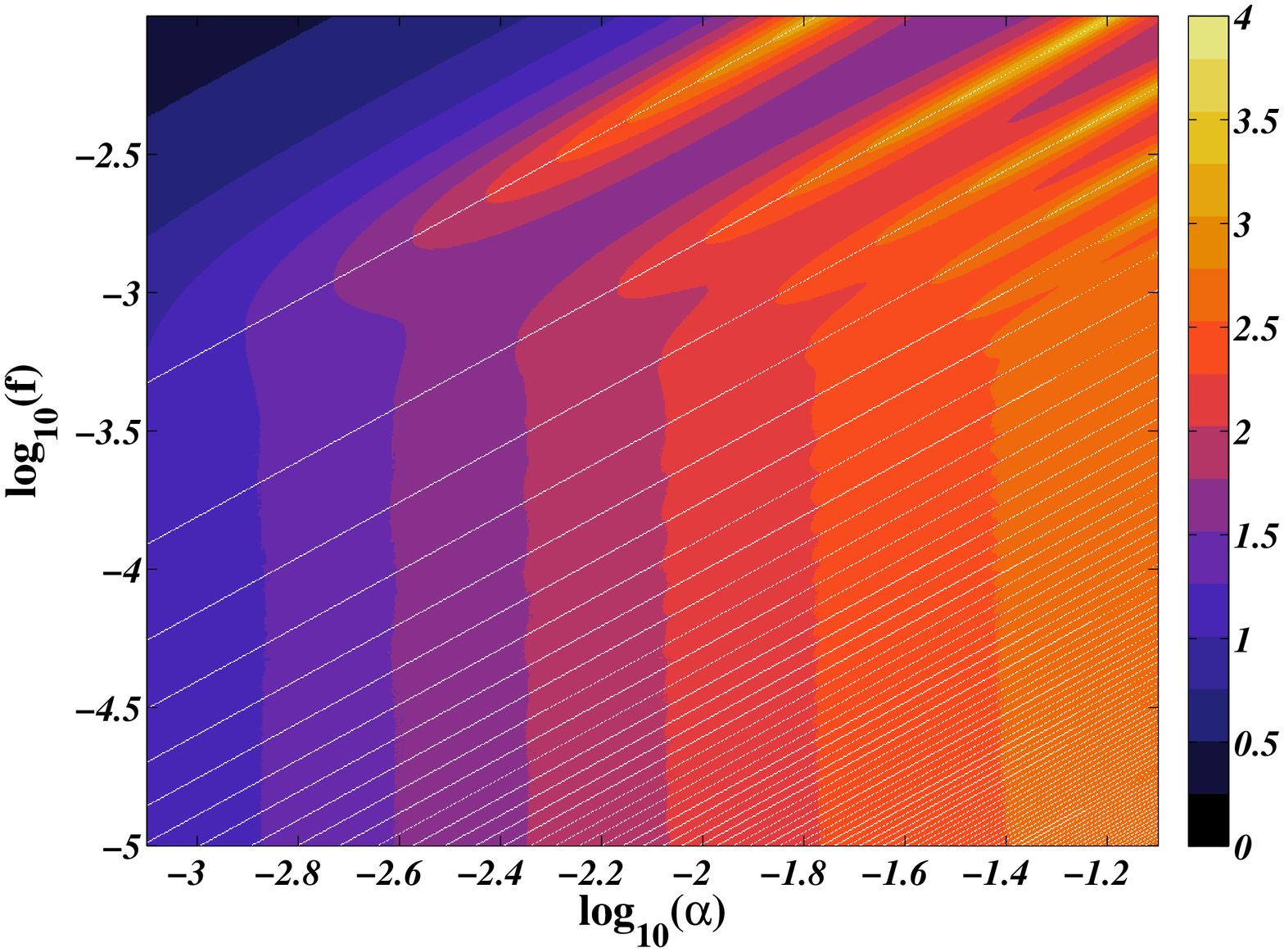}%
\end{center}
\caption{Logarithm of the annihilation cross section in units of $10^{-26}\centi\meter\cubed\per\second$ in a halo with velocity dispersion of approximately $150 \kilo\meter\per\second$. White points are ruled out by either $\mu$- or $|y|$-bound.}%
\label{fig:haloann2_8_200}%
\end{figure}

\begin{figure}
\begin{center}
\includegraphics[height=1.0\columnwidth,angle=270,clip=true]{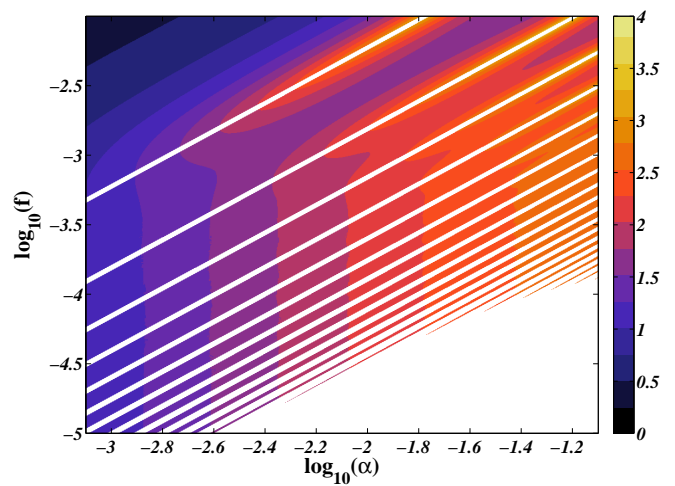}%
\end{center}
\caption{Logarithm of the annihilation cross section in units of $10^{-26}\centi\meter\cubed\per\second$ in a halo with velocity dispersion of approximately $150 \kilo\meter\per\second$. White points can be ruled out if the bound gets improved by a factor of $500$.}%
\label{fig:haloann3_8_200}%
\end{figure}

We can consider what would be the allowed possibilities for the annihilation cross section in a halo having a fiducial velocity dispersion of $150\kilo\meter\per\second$ by removing all points that exceeds either bound. We have plotted this in figure \ref{fig:haloann2_8_200} for our $200\GeV$ example WIMP.
It is suspected that a new FIRAS-like satellite, if built, could bring the bound on $|y|$ down to the order $\sim 10^{-7}$. In figure \ref{fig:haloann3_8_200} we have shown what figure \ref{fig:haloann2_8_200} would look like with a fiducial bound of $|y|<10^{-7}$.
It is also worth comparing this bound with the anisotropy bound. From~\cite{Slatyer:2009yq}, we have the bound
\begin{align}
\tavg{\sigma_\text{ann} v}_\text{saturated} &< \frac{360 \cdot 10^{-26} \centi\meter\cubed\per\second}{F_\text{rc}} \frac{m_\chi}{1 \text{TeV}},
\end{align}
where $F_\text{rc}$ is the average fraction of energy being transfered to the CMB at the time of recombination. This has different values depending on the annihilation channel, as can be seen in table 1 of~\cite{Slatyer:2009yq}. However, it is roughly of order 30\% for most processes.

\begin{figure}
\begin{center}
\includegraphics[height=1.0\columnwidth,angle=270,clip=true]{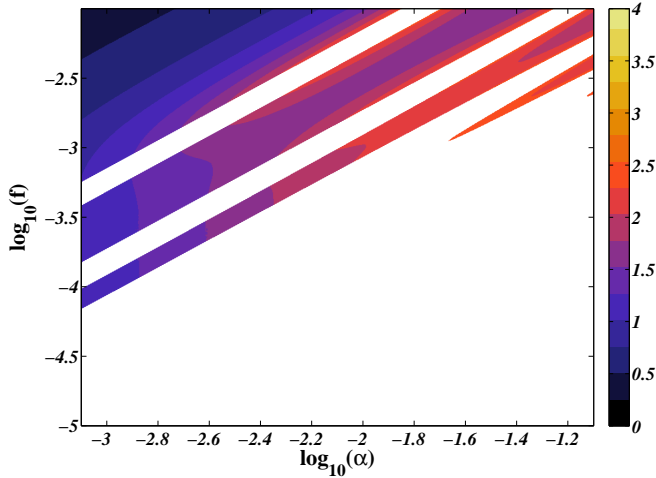}%
\end{center}
\caption{Logarithm of the annihilation cross section in units of $10^{-26}\centi\meter\cubed\per\second$ in a halo with velocity dispersion of approximately $150 \kilo\meter\per\second$. White points are ruled out by the CMB anisotropy bound.}%
\label{fig:haloann4_8_200}%
\end{figure}

As we discussed earlier, plugging in the standard value $\tavg{\sigma_\text{ann} v}_\text{TH}$ for a thermal relic is not strictly accurate. It is easily fixed however, by using our calculated values of $\sigma_0$, as well as the thermally averaged Sommerfeld factor at recombination. We get
\begin{align}
\frac{\sigma_0}{10^{-26}\centi\meter\cubed\per\second} S_\text{avg}(z_\text{rc},\ldots) &<
	\frac{360}{F_\text{rc}} \frac{m_\chi}{1 \text{TeV}}.\label{eq:cmbbound}
\end{align}
In figure~\ref{fig:haloann4_8_200} we have again showed the effective boost factor, but this time with the CMB bound, equation~\eqref{eq:cmbbound}. It is clear, that this bound is stronger than even the forecasted $y$-bound by roughly an order of magnitude.

\section{Conclusion}
We have analysed the Sommerfeld enhancement mechanism in detail and found coordinates in which the position of resonances depends only on one coordinate, $u$, and we have derived analytical expressions for the position of resonances for two model potentials. A numeric treatment of the Sommerfeld case showed that the a similar simple relationship exists for this potential, and we found the following fit to agree excellently:
\begin{equation}
u_n = 0.1592\left(n+1.006\right)^2\qquad n=0,1,\ldots \label{eq:unyukawa}
\end{equation}
This is a nice result, since knowing the position of resonances beforehand is helpful for doing numerical calculations.

In the second part we did the full freeze out calculation for the Sommerfeld parameter space by solving the Boltzmann equation all the way through chemical and kinetic freeze out, and we then calculated the effect of annihilations on the CMB blackbody spectrum. We found that only a small portion of the parameter space, directly on the resonances, can be ruled out by the current bound from FIRAS. We also noted that a future measurement of the $|y|-$ or $\mu-$distortion would rule out a huge portion of the parameter space. But as we showed, the anisotropy bound is already stronger than this and is likely to improve with Planck data. However, we think that it is worth mentioning, that the $|y|-$ and $\mu-$bounds are obtained at a different epoch and by different observations than the anisotropy bounds, and thus should be considered complementary to those.

\appendix
\section{Sommerfeld enhancement for the Coulomb potential}\label{sec:coulomb}
We want to derive equation \eqref{eq:coulombboost}, the boost factor for the Coulomb potential. We start from the radial equation \eqref{eq:radeq1} with $m_\phi=0$ and do the substitution $x = \alpha m_\chi r$:
\begin{align}
\frac{1}{m_\chi} \difff{\chi}{r} &=\left( -\frac{\alpha}{r} e^{-m_\phi r} - m_\chi \beta^2\right) \chi \nonumber \Rightarrow \\
\difff{\chi}{x} &= - \left[ \frac{1}{x} + \left(\frac{\beta}{\alpha} \right)^2 \right] \chi \Rightarrow \nonumber \\
&=\left[ \frac{1}{x} + \epsilon_C^2 \right] \chi. \label{eq:radeqc1}
\end{align}
We analyze equation \eqref{eq:radeqc1} the usual way by considering the asymptotic limits of the equation. We got:
\begin{subequations}
\begin{align}
\difff{\chi}{x} &\simeq -\frac{1}{x} \chi,\phantom{\epsilon_C \chi} x \rightarrow 0 \label{eq:radeqaym1}\\
\difff{\chi}{x} &\simeq -\epsilon_C \chi,\phantom{\frac{1}{x} \chi} x \rightarrow \infty. \label{eq:radeqaym2}
\end{align}
\end{subequations}

The analytic solution of \eqref{eq:radeqaym2} is just an exponential, while the analytic solution of \eqref{eq:radeqaym2} is more complicated. In general we have
\begin{align}
\chi(x) = C_1 \sqrt(x) J_1(2 \sqrt{x}) + C_2 \sqrt{x} Y_1(2\sqrt{x}) \nonumber,
\end{align}
where $J_1$ and $Y_1$ are the Bessel functions of the first and second order, respectively. However, $\sqrt{x}Y_1(2\sqrt{x})$ is not well behaved for $x\rightarrow 0$, so we are left with the $C_1$ term. Since we are looking at $x \ll 1$, we may use $J_1(z) \sim z/2$. Our Ansatz for the solution then becomes:
\begin{align}
\chi &= x e^{i\epsilon_C x} v(x), \label{eq:ansatz}
\end{align} 
where $v(x)$ interpolates between the asymptotic solutions. Taking the second derivative of the Ansatz \eqref{eq:ansatz} yields
\begin{align*}
\difff{\chi}{x} &= e^{i\epsilon_C x} \left[ (2i\epsilon_C - \epsilon_C^2) v + (2+2i \epsilon_C x) v' + xv'' \right],
\end{align*}
which can be inserted into  equation \eqref{eq:radeqc1}. The prime on $v$ denotes differentiation w.r.t $x$. The result is
\begin{align}
0 &= xv'' + (2+2i \epsilon_C x) v' + (2 i \epsilon_C + 1)v.
\end{align}
We can bring this equation on a more recognizable form by making the substitution $z=-2i \epsilon_C x$. This leads to
\begin{align}
0 = z \difff{v}{z} + (2-z) \diff{v}{z} - \left( 1- \frac{i}{2\epsilon_C} \right)v,
\end{align}
which is the confluent hyper geometric equation. The solution which is regular in $x = z = 0$ is then:
\begin{align}
v(z) = C F\left(1-\frac{i}{2\epsilon_C},2,z\right). \label{eq:sol1} 
\end{align}
We now need to apply the asymptotic boundary condition \eqref{eq:bvright} to our solution, so we need the asymptotic formula for $F(a,b,z)$ in the limit $x \rightarrow \infty$ or, equivalently, $z \rightarrow -i\infty$. To do this, define the following:
\begin{subequations}
\begin{align*}
g(a,b,z) &\equiv \sum_{n=0}^{\infty}{\frac{(a)_n (b)_n}{n!} z^{-n}} \\
(c)_n &= c (c+1) (c+2) \cdots (c+n-1),(c)_0=1 \\
F(a,b,z) &= \frac{\Gamma(b)}{\Gamma(b-a)}g(a,1+a-b,-z) (-z)^{-a} +\nonumber \\
         &\frac{\Gamma(b)}{\Gamma(a)}g(b-a,1-a,z) e^zz^{a-b}.
\end{align*}
\end{subequations}
Since $g(a,b,z)\rightarrow 1$ as $|z|\rightarrow \infty$, $F(a,b,z)$ has the asymptotic form
\begin{align*}
F^\infty(a,b,z) &= \frac{\Gamma(b)}{\Gamma(b-a)} (-z)^{-a} +\frac{\Gamma(b)}{\Gamma(a)} e^zz^{a-b}.
\end{align*}
Now consider the gamma-function $\Gamma(b-a)$ for our values of $a=1-i/(2\epsilon_C)$ and $b=2$:
\begin{align*}
\gammaf{b-a} &= \gammaf{2-\left(1-\frac{i}{2\epsilon_C} \right)} = \gammaf{a^*} = \gammaf{a}^*,
\end{align*}
and if we now define $\eta$ by $\gammaf{a} = \abs{\gammaf{a}} e^{i \eta}$, we find the following:
\begin{subequations}
\begin{align}
\frac{\gammaf{b}}{\gammaf{b-a}} &= \frac{\gammaf{b}}{\abs{\gammaf{a}}} e^{i \eta} \\
\frac{\gammaf{b}}{\gammaf{a}} &= \frac{\gammaf{b}}{\abs{\gammaf{a}}} e^{-i \eta}. 
\end{align}
\end{subequations}
The asymptotic form $F^\infty$ can now be written like:
\begin{equation}
\begin{split}
F^\infty&( a,b,z) = \frac{\gammaf{b}}{\abs{\gammaf{a}}} \left(e^{i\eta}(-z)^{-a} + e^{-i \eta+z} z^{a-b} \right) \Rightarrow \nonumber \\
F^\infty&\left( 1-\frac{i}{2\epsilon_C},2,-2i\epsilon_C x \right)  =\frac{e^{\frac{1}{2}z}e^{i\eta-\frac{\pi}{4\epsilon_C}}}{\gammaf{1-\frac{i}{2\epsilon_C}} \epsilon_C x} \times \nonumber \\
&\times\sin \left( \epsilon_C x + \frac{1}{2\epsilon_C} \ln(2\epsilon_C x) + \eta \right).
\end{split}
\end{equation}
Substituting the asymptotic solution back into the Ansatz \eqref{eq:ansatz} yields
\begin{multline}
\chi^\infty(x) = C_1 \frac{e^{i\eta-\frac{\pi}{4\epsilon_C}}}{\gammaf{1-\frac{i}{2\epsilon_C}} \epsilon_C } \times \\ \times \sin \left( \epsilon_C x + \frac{1}{2\epsilon_C} \ln(2\epsilon_C x) + \eta \right),
\end{multline}
where $C_1$ is now set by the boundary condition \eqref{eq:bvright}. We find
\begin{align}
C_1 &= \frac{\epsilon_C \gammaf{1-\frac{i}{2\epsilon_C}}}{e^{i\eta-\frac{\pi}{4\epsilon_C}}},
\end{align}
which we can finally insert into \eqref{eq:sol1} and the Ansatz \eqref{eq:ansatz}:
\begin{multline}
\chi(x) = x e^{i \epsilon_C x} \frac{\epsilon_C \gammaf{1-\frac{i}{2\epsilon_C}}}{e^{i\eta-\frac{\pi}{4\epsilon_C}}} \times \\ \times F\left(1-\frac{i}{2\epsilon_C},2,-2 i \epsilon_C x \right). \label{eq:sol2}
\end{multline}
We are interested in the boost factor~\eqref{eq:boostcalc}, so we calculate
\begin{align}
S_{\epsilon_C} &= \abs{\frac{R_{k0}}{k}}^2 = \abs{\frac{1}{\epsilon_C} \diff{\chi}{x}(x=0)}^2.
\end{align}
We take the derivative of our solution \eqref{eq:sol2} and evaluate it in $x=0$:
\begin{align}
\diff{\chi}{x}(x=0) &=\epsilon_C \gammaf{1-\frac{i}{2\epsilon_C}} e^{-i\eta +\frac{\pi}{4\epsilon_C}} \Rightarrow \nonumber\\
S_{\epsilon_C} &=\abs{\gammaf{1-\frac{i}{2\epsilon_C}} e^{-i\eta +\frac{\pi}{4\epsilon_C}}}^2 \nonumber \\
        &=\abs{\gammaf{1-\frac{i}{2\epsilon_C}}}^2 e^{\frac{\pi}{2\epsilon_C}} \nonumber \\
	&=\frac{-\frac{\pi}{2\epsilon_C}}{\sinh\left(-\frac{\pi}{2\epsilon_C}\right)} e^{\frac{\pi}{2\epsilon_C}} \nonumber \\
	&= \frac{\pi/\epsilon_C}{1-e^{-\pi/\epsilon_C}},
\end{align}
which is the result obtained by Sommerfeld and quoted in~\cite{ArkaniHamed:2008qn}.

\section{How to solve the integrated Boltzmann equation}\label{sec:boltzmann}
We want to derive the evolution equation \eqref{eq:evoboltz} for the number density of Dark Matter. Starting from the integrated Boltzmann equation, we get:
\begin{align}
a^{-3} \diff{(n_\chi a^3)}{t} &= \tavg{\sigma_\text{ann}v} \left\{ n_{\chi,\text{eq}}^2-n_\chi^2 \right\},
\end{align}
where $a$ is the scale factor, $n_\chi$ is the number density of Dark Matter, $n_{\chi,\text{eq}}$ is the Dark Matter equilibrium number density and $\tavg{\sigma_\text{ann}v}$ is the thermally averaged annihilation cross section. We normalize the number density to the total entropy density $s\propto a^{-3}$ by introducing $Y\equiv \frac{n_\chi}{s}$:
\begin{equation}
\diff{Y}{t} = s \tavg{\sigma_\text{ann} v} \left\{ Y_\text{eq}^2 - Y^2 \right\}.
\end{equation}
We want to substitute the time parameter $t$ by the dimensionless evolution parameter $x\equiv \frac{m_\chi}{T_\gamma}$. We need a few formulas to proceed. Using $s\propto a^{-3}$, we find:
\begin{equation}
\diff{s a^3}{t}=0 \Rightarrow \dot{s} = -3sH. \label{eq:sdotsH}
\end{equation}
The entropy density $s$ is given by
\begin{align}
s(T) &= k g_{*S}(T) T^3,
\end{align}
where $k=\frac{2\pi^2}{45}$ is a constant and $g_{*S}$ is the relativistic degrees of freedom for the entropy density. Taking the time derivative yields:
\begin{align}
\dot{s} &=\diff{s}{T} \dot{T} =\dot{T} \left[ k \diff{g_{*S}}{T} T^3+3kg_{*S} T^3 \frac{1}{T} \right] \nonumber\\
				&=3s\frac{\dot{T}}{T} \left[\diff{g_{*S}}{T} \frac{1}{3g_{*S}} + 1 \right] \equiv 3s\frac{\dot{T}}{T} \frac{g_*^\frac{1}{2} h_*^\frac{1}{2}}{g_{*S}}, \label{eq:hstar}
\end{align}
where the last line defines the parameter $h_*^\frac{1}{2}$. Equating equation \eqref{eq:sdotsH} and \eqref{eq:hstar}, yields
\begin{align}
\frac{\dot{T}}{T} &= -H \frac{g_{*S}}{g_*^\frac{1}{2} h_*^\frac{1}{2}} = - \sqrt{4\pi^3}{45} T^2 g_{*S}(T) \frac{1}{m_p h_*^\frac{1}{2}},
\end{align}
where the Friedmann equation in the radiation dominated universe has been used in the last equation. Using the chain rule on the left hand side of \eqref{eq:boltzmann2} with this equation gives
\begin{align}
\diff{Y}{t} &= \diff{Y}{x} \diff{x}{T} \diff{T}{t} = \diff{Y}{x} x \left( \frac{\dot{T}}{T} \right) \nonumber\\
						&= \diff{Y}{x} x \sqrt{\frac{4\pi^3}{45}} \left(\frac{m_\chi}{x} \right)^2 g_{*S}(T) \frac{1}{m_p h_*^\frac{1}{2}}.
\end{align}
This gives our final form of the evolution equation:
\begin{align}
\diff{Y}{x} &= \sqrt{\frac{\pi}{45}} \frac{m_p m_\chi}{x^2}h_*^\frac{1}{2}\tavg{\sigma_\text{ann} v} \left\{ Y_\text{eq}^2 - Y^2 \right\}.
\end{align}

\subsection{Calculating the average number density}
The number density of Dark Matter in equilibrium, $n_\text{eq}$, can be found by integrating the distribution function $f(\vect{x},\vect{p})$ over momentum space. We assume a Maxwell-Boltzmann distribution with zero chemical potential:
\begin{align}
f(\vect{x},\vect{p}) &= g e^{\frac{-E(p)}{T}}.
\end{align}
Using $E(p) = \sqrt{p^2 + m^2}$, and introducing $x=m_\chi/T$ and $\alpha=p/T$ we get:
\begin{align}
n_\text{eq} &= g \int \frac{\mathrm{d}^3p}{(2\pi)^3} e^{-\frac{E}{T}} \nonumber \\
						&= \frac{4\pi g T^3}{(2\pi)^3}  \int_1^\infty \mathrm{d}\alpha \alpha^2 e^{-\sqrt{\alpha^2+x^2}}\nonumber \\
						&= \frac{4\pi g T^3}{(2\pi)^3} x^3 \int_1^\infty \mathrm{d}t t \sqrt{t^2-1} e^{-xt} \nonumber\\
						& =\frac{4\pi g T^3}{(2\pi)^3} x^2 K_2(x) =\frac{g m_\chi^3}{2\pi^2 x} K_2(x) \Rightarrow \nonumber\\
Y_\text{eq}	& = g \frac{45}{4\pi^4} \frac{x^2}{g_{*S}} K_2(x),
\end{align}
where $K_2$ is the modified Bessel function of the second kind.

\subsection{Solving the differential equation}
Equation \eqref{eq:boltzmann3} is an example of a stiff differential equation, meaning that it involves vastly different scales. In this case it is roughly the timescale at which annihilations take place versus the timescale at which expansion happens. Explicit solvers like Runge-Kutta either fails or need a very low step size to maintain stability. We implement the implicit scheme, which is also used in the software package DarkSUSY~\cite{DarkSUSY,Gondolo:2004sc}. The derivative $y'$ of a function can be approximated as follows:
\begin{subequations}
\label{eq:eulerdiff}
\begin{align} 
y'_i			&\simeq \frac{y_{i+1}-y_i}{h} \qquad \text{Forward difference} \\
y'_{i+1} 	&\simeq \frac{y_{i+1}-y_i}{h} \qquad \text{Backward difference}.\label{eq:backdiff}
\end{align}
\end{subequations}
Equation \eqref{eq:backdiff} inserted in the general ODE $y'=f(x,y)$ gives rise to the implicit backward euler scheme:
\begin{align}
y_{i+1} &= y_i + h f_{i+1}(x,y) + \bigo{h}.\label{eq:bweuler}
\end{align}
Adding equations \eqref{eq:eulerdiff} yields the implicit trapezoidal rule
\begin{align}
y_{i+1} &= y_i + \frac{h}{2} \left( f_{i+1} + f_i \right) + \bigo{h^2}. \label{eq:trapezoidal}
\end{align}
These two schemes are the $s=0$ and $s=1$ Adams-Moulton methods respectively. We can use the difference between the two methods as an error-estimate for controlling the step size. An implicit scheme like this usually results in a system of non-linear algebraic equations which must then be solved numerically in each step. Fortunately, for the Boltzmann-equation, and using method \eqref{eq:bweuler} or \eqref{eq:trapezoidal}, this is just a second order equation which can be solved analytically. We introduce the variables suggested in~\cite{DarkSUSY}:
\begin{subequations}
\begin{align}
Y_{eq} 	&\equiv q \\
u				&\equiv h \lambda_{i+1} \\
\rho		&\equiv \frac{\lambda_i}{\lambda_{i+1}} \\
c				&\equiv 2Y_i + u \left[(q_{i+1}^2+\rho q_i^2) -\rho Y_i^2 \right] \\
c'			&\equiv 4(Y_i + u q_{i+1}^2).
\end{align}
\end{subequations}
The Euler method \eqref{eq:bweuler} for the Boltzmann equation \eqref{eq:boltzmann3} is
\begin{align}
4 Y_{i+1} &=4 y_i + 4h\lambda_{i+1}(q_{i+1}^2-Y_{i+1}^2) \nonumber\\
					&=c' - 4uy_{i+1}^2 \Rightarrow \nonumber\\
Y_{i+1}		&=-\frac{1 \mp \sqrt{1+uc'}}{2u} \Rightarrow \nonumber\\
Y_{i+1}		&=\frac{1}{2} \frac{c'}{1 + \sqrt{1 + u c'}},
\end{align}
where we have chosen the solution which gives a positive $Y_{i+1}$. It works the same way for the trapezoidal rule \eqref{eq:trapezoidal}:
\begin{align}
2 Y_{i+1} &=2Y_i + h\left[ \lambda_i(q_i^2 - Y_i^2) + \lambda_{i+1}(q_{i+1}^2-Y_{i+1}^2) 									\right]\nonumber\\
					&=2Y_i + u \left[\rho q_i^2 + q_{i+1}^2 - \rho Y_i^2\right] - uY_{i+1}^2 \nonumber\\
 					&=c - u Y_{i+1}^2 \Rightarrow \nonumber\\
Y_{i+1}		&=\frac{-1 \pm \sqrt{1+uc}{u}}{u} \Rightarrow\nonumber\\
Y_{i+1}		&=\frac{c}{1 + \sqrt{1 + uc}}.
\end{align}
If we let $Y'_{i+1}$ denote the Euler estimate of the next step, we estimate the relative error $\text{err}$ as
\begin{align}
\text{err} &= \left| \frac{Y_{i+1}'-Y_{i+1}}{Y_{i+1}} \right|,
\end{align}
and since the trapezoidal rule is second order in h, we modify the step size according to
\begin{align}
\text{hnext} &= \min \left(hS\sqrt{\frac{\text{eps}}{\text{err}}},5h \right),
\end{align}
where $S$ is a safety factor set at $0.9$ and $\text{eps}$ is the wanted accuracy. We have also demanded that $h$ can only grow with a factor of $5$ in each step.
\bibliographystyle{../../../LatexStuff/bibstyles/utcaps}
\bibliography{../../../LatexStuff/references/cmbbound}

\providecommand{\href}[2]{#2}\begingroup\raggedright\begin{thebibliography}{10}

\bibitem{Adriani:2008zr}
{\bfseries PAMELA} Collaboration, O.~Adriani {\em et al.}, ``{An anomalous
  positron abundance in cosmic rays with energies 1.5.100 GeV},''
  \href{http://dx.doi.org/10.1038/nature07942}{{\em Nature} {\bfseries 458}
  (2009)  607--609},
\href{http://arxiv.org/abs/0810.4995}{{\ttfamily arXiv:0810.4995 [astro-ph]}}.
%%CITATION = 0810.4995;%%.

\bibitem{Barwick:1997ig}
{\bfseries HEAT} Collaboration, S.~W. Barwick {\em et al.}, ``{Measurements of
  the cosmic-ray positron fraction from 1- GeV to 50-GeV},'' {\em Astrophys.
  J.} {\bfseries 482} (1997)  L191--L194,
\href{http://arxiv.org/abs/astro-ph/9703192}{{\ttfamily
  arXiv:astro-ph/9703192}}.
%%CITATION = ASTRO-PH/9703192;%%.

\bibitem{Beatty:2004cy}
J.~J. Beatty {\em et al.}, ``{New measurement of the cosmic-ray positron
  fraction from 5-GeV to 15-GeV},''
  \href{http://dx.doi.org/10.1103/PhysRevLett.93.241102}{{\em Phys. Rev. Lett.}
  {\bfseries 93} (2004)  241102},
\href{http://arxiv.org/abs/astro-ph/0412230}{{\ttfamily
  arXiv:astro-ph/0412230}}.
%%CITATION = ASTRO-PH/0412230;%%.

\bibitem{Aguilar:2007yf}
{\bfseries AMS-01} Collaboration, M.~Aguilar {\em et al.}, ``{Cosmic-ray
  positron fraction measurement from 1-GeV to 30- GeV with AMS-01},''
  \href{http://dx.doi.org/10.1016/j.physletb.2007.01.024}{{\em Phys. Lett.}
  {\bfseries B646} (2007)  145--154},
\href{http://arxiv.org/abs/astro-ph/0703154}{{\ttfamily
  arXiv:astro-ph/0703154}}.
%%CITATION = ASTRO-PH/0703154;%%.

\bibitem{Chang:2008zzr}
J.~Chang {\em et al.}, ``{An excess of cosmic ray electrons at energies of
  300.800 GeV},''
\href{http://dx.doi.org/10.1038/nature07477}{{\em Nature} {\bfseries 456}
  (2008)  362--365}.
%%CITATION = NATUA,456,362;%%.

\bibitem{Finkbeiner:2003im}
D.~P. Finkbeiner, ``{Microwave ISM Emission Observed by WMAP},''
  \href{http://dx.doi.org/10.1086/423482}{{\em Astrophys. J.} {\bfseries 614}
  (2004)  186--193},
\href{http://arxiv.org/abs/astro-ph/0311547}{{\ttfamily
  arXiv:astro-ph/0311547}}.
%%CITATION = ASTRO-PH/0311547;%%.

\bibitem{Dobler:2007wv}
G.~Dobler and D.~P. Finkbeiner, ``{Extended Anomalous Foreground Emission in
  the WMAP 3-Year Data},'' \href{http://dx.doi.org/10.1086/587862}{{\em
  Astrophys. J.} {\bfseries 680} (2008)  1222--1234},
\href{http://arxiv.org/abs/0712.1038}{{\ttfamily arXiv:0712.1038 [astro-ph]}}.
%%CITATION = 0712.1038;%%.

\bibitem{Strong:2005zx}
A.~W. Strong {\em et al.}, ``{Gamma-ray continuum emission from the inner
  Galactic region as observed with INTEGRAL/SPI},''
  \href{http://dx.doi.org/10.1051/0004-6361:20053798}{{\em Astron. Astrophys.}
  {\bfseries 444} (2005)  495},
\href{http://arxiv.org/abs/astro-ph/0509290}{{\ttfamily
  arXiv:astro-ph/0509290}}.
%%CITATION = ASTRO-PH/0509290;%%.

\bibitem{Thompson:2004ez}
D.~J. Thompson, D.~L. Bertsch, and R.~H. O'Neal, Jr., ``{The highest-energy
  photons seen by the Energetic Gamma Ray Experiment Telescope (EGRET) on the
  Compton Gamma Ray Observatory},''
\href{http://arxiv.org/abs/astro-ph/0412376}{{\ttfamily
  arXiv:astro-ph/0412376}}.
%%CITATION = ASTRO-PH/0412376;%%.

\bibitem{Kane:2002nm}
G.~L. Kane, L.-T. Wang, and T.~T. Wang, ``{Supersymmetry and the cosmic ray
  positron excess},''
  \href{http://dx.doi.org/10.1016/S0370-2693(02)01839-7}{{\em Phys. Lett.}
  {\bfseries B536} (2002)  263--269},
\href{http://arxiv.org/abs/hep-ph/0202156}{{\ttfamily arXiv:hep-ph/0202156}}.
%%CITATION = HEP-PH/0202156;%%.

\bibitem{Hooper:2003ad}
D.~Hooper, J.~E. Taylor, and J.~Silk, ``{Can supersymmetry naturally explain
  the positron excess?},''
  \href{http://dx.doi.org/10.1103/PhysRevD.69.103509}{{\em Phys. Rev.}
  {\bfseries D69} (2004)  103509},
\href{http://arxiv.org/abs/hep-ph/0312076}{{\ttfamily arXiv:hep-ph/0312076}}.
%%CITATION = HEP-PH/0312076;%%.

\bibitem{Cholis:2008qq}
I.~Cholis, D.~P. Finkbeiner, L.~Goodenough, and N.~Weiner, ``{The PAMELA
  Positron Excess from Annihilations into a Light Boson},''
  \href{http://dx.doi.org/10.1088/1475-7516/2009/12/007}{{\em JCAP} {\bfseries
  0912} (2009)  007},
\href{http://arxiv.org/abs/0810.5344}{{\ttfamily arXiv:0810.5344 [astro-ph]}}.
%%CITATION = 0810.5344;%%.

\bibitem{Baltz:2001ir}
E.~A. Baltz, J.~Edsjo, K.~Freese, and P.~Gondolo, ``{The cosmic ray positron
  excess and neutralino dark matter},''
  \href{http://dx.doi.org/10.1103/PhysRevD.65.063511}{{\em Phys. Rev.}
  {\bfseries D65} (2002)  063511},
\href{http://arxiv.org/abs/astro-ph/0109318}{{\ttfamily
  arXiv:astro-ph/0109318}}.
%%CITATION = ASTRO-PH/0109318;%%.

\bibitem{Hooper:2007kb}
D.~Hooper, D.~P. Finkbeiner, and G.~Dobler, ``{Evidence Of Dark Matter
  Annihilations In The WMAP Haze},''
  \href{http://dx.doi.org/10.1103/PhysRevD.76.083012}{{\em Phys. Rev.}
  {\bfseries D76} (2007)  083012},
\href{http://arxiv.org/abs/0705.3655}{{\ttfamily arXiv:0705.3655 [astro-ph]}}.
%%CITATION = 0705.3655;%%.

\bibitem{Cirelli:2008pk}
M.~Cirelli, M.~Kadastik, M.~Raidal, and A.~Strumia, ``{Model-independent
  implications of the e+, e-, anti-proton cosmic ray spectra on properties of
  Dark Matter},'' \href{http://dx.doi.org/10.1016/j.nuclphysb.2008.11.031}{{\em
  Nucl. Phys.} {\bfseries B813} (2009)  1--21},
\href{http://arxiv.org/abs/0809.2409}{{\ttfamily arXiv:0809.2409 [hep-ph]}}.
%%CITATION = 0809.2409;%%.

\bibitem{ArkaniHamed:2008qn}
N.~Arkani-Hamed, D.~P. Finkbeiner, T.~R. Slatyer, and N.~Weiner, ``{A Theory of
  Dark Matter},'' \href{http://dx.doi.org/10.1103/PhysRevD.79.015014}{{\em
  Phys. Rev.} {\bfseries D79} (2009)  015014},
\href{http://arxiv.org/abs/0810.0713}{{\ttfamily arXiv:0810.0713 [hep-ph]}}.
%%CITATION = 0810.0713;%%.

\bibitem{Hisano:2003ec}
J.~Hisano, S.~Matsumoto, and M.~M. Nojiri, ``{Explosive dark matter
  annihilation},'' \href{http://dx.doi.org/10.1103/PhysRevLett.92.031303}{{\em
  Phys. Rev. Lett.} {\bfseries 92} (2004)  031303},
\href{http://arxiv.org/abs/hep-ph/0307216}{{\ttfamily arXiv:hep-ph/0307216}}.
%%CITATION = HEP-PH/0307216;%%.

\bibitem{Hisano:2004ds}
J.~Hisano, S.~Matsumoto, M.~M. Nojiri, and O.~Saito, ``{Non-perturbative effect
  on dark matter annihilation and gamma ray signature from galactic center},''
  \href{http://dx.doi.org/10.1103/PhysRevD.71.063528}{{\em Phys. Rev.}
  {\bfseries D71} (2005)  063528},
\href{http://arxiv.org/abs/hep-ph/0412403}{{\ttfamily arXiv:hep-ph/0412403}}.
%%CITATION = HEP-PH/0412403;%%.

\bibitem{Cirelli:2007xd}
M.~Cirelli, A.~Strumia, and M.~Tamburini, ``{Cosmology and Astrophysics of
  Minimal Dark Matter},''
  \href{http://dx.doi.org/10.1016/j.nuclphysb.2007.07.023}{{\em Nucl. Phys.}
  {\bfseries B787} (2007)  152--175},
\href{http://arxiv.org/abs/0706.4071}{{\ttfamily arXiv:0706.4071 [hep-ph]}}.
%%CITATION = 0706.4071;%%.

\bibitem{MarchRussell:2008yu}
J.~March-Russell, S.~M. West, D.~Cumberbatch, and D.~Hooper, ``{Heavy Dark
  Matter Through the Higgs Portal},''
  \href{http://dx.doi.org/10.1088/1126-6708/2008/07/058}{{\em JHEP} {\bfseries
  07} (2008)  058},
\href{http://arxiv.org/abs/0801.3440}{{\ttfamily arXiv:0801.3440 [hep-ph]}}.
%%CITATION = 0801.3440;%%.

\bibitem{Cassel:2009wt}
S.~Cassel, ``{Sommerfeld factor for arbitrary partial wave processes},''
  \href{http://dx.doi.org/10.1088/0954-3899/37/10/105009}{{\em J. Phys.}
  {\bfseries G37} (2010)  105009},
\href{http://arxiv.org/abs/0903.5307}{{\ttfamily arXiv:0903.5307 [hep-ph]}}.
%%CITATION = 0903.5307;%%.

\bibitem{Slatyer:2009vg}
T.~R. Slatyer, ``{The Sommerfeld enhancement for dark matter with an excited
  state},'' \href{http://dx.doi.org/10.1088/1475-7516/2010/02/028}{{\em JCAP}
  {\bfseries 1002} (2010)  028},
\href{http://arxiv.org/abs/0910.5713}{{\ttfamily arXiv:0910.5713 [hep-ph]}}.
%%CITATION = 0910.5713;%%.

\bibitem{Cirelli:2009bb}
M.~Cirelli, F.~Iocco, and P.~Panci, ``{Constraints on Dark Matter annihilations
  from reionization and heating of the intergalactic gas},''
  \href{http://dx.doi.org/10.1088/1475-7516/2009/10/009}{{\em JCAP} {\bfseries
  0910} (2009)  009},
\href{http://arxiv.org/abs/0907.0719}{{\ttfamily arXiv:0907.0719
  [astro-ph.CO]}}.
%%CITATION = 0907.0719;%%.

\bibitem{Chen:2003gz}
X.-L. Chen and M.~Kamionkowski, ``{Particle decays during the cosmic dark
  ages},'' \href{http://dx.doi.org/10.1103/PhysRevD.70.043502}{{\em Phys. Rev.}
  {\bfseries D70} (2004)  043502},
\href{http://arxiv.org/abs/astro-ph/0310473}{{\ttfamily
  arXiv:astro-ph/0310473}}.
%%CITATION = ASTRO-PH/0310473;%%.

\bibitem{Pierpaoli:2003rz}
E.~Pierpaoli, ``{Decaying particles and the reionization history of the
  universe},'' \href{http://dx.doi.org/10.1103/PhysRevLett.92.031301}{{\em
  Phys. Rev. Lett.} {\bfseries 92} (2004)  031301},
\href{http://arxiv.org/abs/astro-ph/0310375}{{\ttfamily
  arXiv:astro-ph/0310375}}.
%%CITATION = ASTRO-PH/0310375;%%.

\bibitem{Furlanetto:2006wp}
S.~R. Furlanetto, S.~P. Oh, and E.~Pierpaoli, ``{The Effects of Dark Matter
  Decay and Annihilation on the High-Redshift 21 cm Background},''
  \href{http://dx.doi.org/10.1103/PhysRevD.74.103502}{{\em Phys. Rev.}
  {\bfseries D74} (2006)  103502},
\href{http://arxiv.org/abs/astro-ph/0608385}{{\ttfamily
  arXiv:astro-ph/0608385}}.
%%CITATION = ASTRO-PH/0608385;%%.

\bibitem{Mapelli:2006ej}
M.~Mapelli, A.~Ferrara, and E.~Pierpaoli, ``{Impact of dark matter decays and
  annihilations on reionization},''
  \href{http://dx.doi.org/10.1111/j.1365-2966.2006.10408.x}{{\em Mon. Not. Roy.
  Astron. Soc.} {\bfseries 369} (2006)  1719--1724},
\href{http://arxiv.org/abs/astro-ph/0603237}{{\ttfamily
  arXiv:astro-ph/0603237}}.
%%CITATION = ASTRO-PH/0603237;%%.

\bibitem{Zhang:2007zzh}
L.~Zhang, X.~Chen, M.~Kamionkowski, Z.-g. Si, and Z.~Zheng, ``{Contraints on
  radiative dark-matter decay from the cosmic microwave background},''
  \href{http://dx.doi.org/10.1103/PhysRevD.76.061301}{{\em Phys. Rev.}
  {\bfseries D76} (2007)  061301},
\href{http://arxiv.org/abs/0704.2444}{{\ttfamily arXiv:0704.2444 [astro-ph]}}.
%%CITATION = 0704.2444;%%.

\bibitem{Galli:2009zc}
S.~Galli, F.~Iocco, G.~Bertone, and A.~Melchiorri, ``{CMB constraints on Dark
  Matter models with large annihilation cross-section},''
  \href{http://dx.doi.org/10.1103/PhysRevD.80.023505}{{\em Phys. Rev.}
  {\bfseries D80} (2009)  023505},
\href{http://arxiv.org/abs/0905.0003}{{\ttfamily arXiv:0905.0003
  [astro-ph.CO]}}.
%%CITATION = 0905.0003;%%.

\bibitem{Slatyer:2009yq}
T.~R. Slatyer, N.~Padmanabhan, and D.~P. Finkbeiner, ``{CMB Constraints on WIMP
  Annihilation: Energy Absorption During the Recombination Epoch},''
  \href{http://dx.doi.org/10.1103/PhysRevD.80.043526}{{\em Phys. Rev.}
  {\bfseries D80} (2009)  043526},
\href{http://arxiv.org/abs/0906.1197}{{\ttfamily arXiv:0906.1197
  [astro-ph.CO]}}.
%%CITATION = 0906.1197;%%.

\bibitem{Dent:2009bv}
J.~B. Dent, S.~Dutta, and R.~J. Scherrer, ``{Thermal Relic Abundances of
  Particles with Velocity- Dependent Interactions},''
\href{http://arxiv.org/abs/0909.4128}{{\ttfamily arXiv:0909.4128
  [astro-ph.CO]}}.
%%CITATION = 0909.4128;%%.

\bibitem{Feng:2009hw}
J.~L. Feng, M.~Kaplinghat, and H.-B. Yu, ``{Halo Shape and Relic Density
  Exclusions of Sommerfeld- Enhanced Dark Matter Explanations of Cosmic Ray
  Excesses},'' \href{http://dx.doi.org/10.1103/PhysRevLett.104.151301}{{\em
  Phys. Rev. Lett.} {\bfseries 104} (2010)  151301},
\href{http://arxiv.org/abs/0911.0422}{{\ttfamily arXiv:0911.0422 [hep-ph]}}.
%%CITATION = 0911.0422;%%.

\bibitem{Feng:2010zp}
J.~L. Feng, M.~Kaplinghat, and H.-B. Yu, ``{Sommerfeld Enhancements for Thermal
  Relic Dark Matter},''
\href{http://arxiv.org/abs/1005.4678}{{\ttfamily arXiv:1005.4678 [hep-ph]}}.
%%CITATION = 1005.4678;%%.

\bibitem{DarkSUSY}
P.~Gondolo, J.~Edsjö, P.~Ullio, L.~Bergström, M.~Schelke, E.~Baltz,
  T.~Bringmann, and G.~Duda, ``{DarkSUSY homepage}.'' Online.

\bibitem{Bernstein:1989uq}
J.~Bernstein and S.~Dodelson, ``{ASPECTS OF THE ZELDOVICH-SUNYAEV MECHANISM},''
\href{http://dx.doi.org/10.1103/PhysRevD.41.354}{{\em Phys. Rev.} {\bfseries
  D41} (1990)  354}.
%%CITATION = PHRVA,D41,354;%%.

\bibitem{McDonald:2000bk}
P.~McDonald, R.~J. Scherrer, and T.~P. Walker, ``{Cosmic microwave background
  constraint on residual annihilations of relic particles},''
  \href{http://dx.doi.org/10.1103/PhysRevD.63.023001}{{\em Phys. Rev.}
  {\bfseries D63} (2001)  023001},
\href{http://arxiv.org/abs/astro-ph/0008134}{{\ttfamily
  arXiv:astro-ph/0008134}}.
%%CITATION = ASTRO-PH/0008134;%%.

\bibitem{Zavala:2009mi}
J.~Zavala, M.~Vogelsberger, and S.~D.~M. White, ``{Relic density and CMB
  constraints on dark matter annihilation with Sommerfeld enhancement},''
\href{http://arxiv.org/abs/0910.5221}{{\ttfamily arXiv:0910.5221
  [astro-ph.CO]}}.
%%CITATION = 0910.5221;%%.

\bibitem{Bringmann:2009vf}
T.~Bringmann, ``{Particle Models and the Small-Scale Structure of Dark
  Matter},'' \href{http://dx.doi.org/10.1088/1367-2630/11/10/105027}{{\em New
  J. Phys.} {\bfseries 11} (2009)  105027},
\href{http://arxiv.org/abs/0903.0189}{{\ttfamily arXiv:0903.0189
  [astro-ph.CO]}}.
%%CITATION = 0903.0189;%%.

\bibitem{Fixsen:1996nj}
D.~J. Fixsen {\em et al.}, ``{The Cosmic Microwave Background Spectrum from the
  Full COBE/FIRAS Data Set},'' \href{http://dx.doi.org/10.1086/178173}{{\em
  Astrophys. J.} {\bfseries 473} (1996)  576},
\href{http://arxiv.org/abs/astro-ph/9605054}{{\ttfamily
  arXiv:astro-ph/9605054}}.
%%CITATION = ASTRO-PH/9605054;%%.

\bibitem{Gondolo:2004sc}
P.~Gondolo {\em et al.}, ``{DarkSUSY: Computing supersymmetric dark matter
  properties numerically},''
  \href{http://dx.doi.org/10.1088/1475-7516/2004/07/008}{{\em JCAP} {\bfseries
  0407} (2004)  008},
\href{http://arxiv.org/abs/astro-ph/0406204}{{\ttfamily
  arXiv:astro-ph/0406204}}.
%%CITATION = ASTRO-PH/0406204;%%.

\end{thebibliography}\endgroup
\end{document}